\documentclass[a4paper,fleqn]{cas-dc}

\usepackage[authoryear]{natbib}

\usepackage{amssymb}
\usepackage[utf8]{inputenc}
\usepackage{amsmath}
\usepackage{xcolor}
\usepackage[normalem]{ulem}
\usepackage{graphicx}
\usepackage{bm}
\usepackage{stfloats}

\def\tsc#1{\csdef{#1}{\textsc{\lowercase{#1}}\xspace}}
\tsc{VP}
\tsc{NT}
\tsc{EC}

\newcommand{\der}[2]{\frac{\mathrm{d} #1}{\mathrm{d} #2}}
\newcommand{\vt}{v_\mathrm{t}}
\newcommand{\avt}{|v_\mathrm{t}|}
\newcommand{\vm}{u_\mathrm{rms}}
\newcommand{\vsh}{v_\mathrm{sh}}
\newcommand{\rc}{r_\mathrm{c}}

\newcommand{\St}{St}

\newcommand{\tr}{t_\mathrm{res}}
\newcommand{\mrt}{\overline{t_\mathrm{res}}}

\begin{document}
\let\WriteBookmarks\relax
\def\floatpagepagefraction{1}
\def\textpagefraction{.001}
\shorttitle{Residence Time of Inertial Particles in Convecting Fluids}
\shortauthors{Pato\v{c}ka et al.}

\title [mode = title]{Residence time of inertial particles in 3D thermal convection: implications for magma reservoirs.}           

\author[1]{Vojt\v{e}ch Pato\v{c}ka}[orcid=0000-0002-3413-6120]
\credit{Conceptualization, Methodology, Investigation, Visualization, Writing - Original draft preparation}
\address[1]{Faculty of Mathematics and Physics, Department of Geophysics, Charles University, Prague, Czech Republic}

\author[2]{Nicola Tosi}[]
\credit{Conceptualization, Methodology, Writing - Review \& Editing, Supervision}
\address[2]{Institute for Planetary Research, German Aerospace Center (DLR), Berlin, Germany}

\author[3]{Enrico Calzavarini}[]
\credit{Conceptualization, Methodology, Software, Writing - Review \& Editing, Supervision}
\address[3]{Univ. Lille, ULR 7512 - Unité de Mécanique de Lille - Joseph Boussinesq (UML), F-59000 Lille, France}

\begin{abstract}
The dynamic behaviour of crystals in convecting fluids determines how magma bodies solidify. In particular, it is often important to estimate how long crystals stay in suspension in the host liquid before being deposited at its bottom (or top, for light crystals and bubbles of volatiles). We perform a systematic 3D numerical study of particle-laden Rayleigh-B\'{e}nard convection, and derive a robust model for the particle residence time. For Rayleigh numbers higher than $10^7$, inertial particles' trajectories exhibit a monotonic transition from fluid tracer-like to free-fall dynamics, the control parameter being the ratio between particle Stokes velocity and the mean amplitude of the fluid velocity. The average settling rate is proportional to the particle Stokes velocity in both the end-member regimes, but the distribution of the residence times differs markedly from one to the other. For lower Rayleigh numbers ($<10^7$), an interaction between large-scale circulation and particle motion emerges, increasing the settling rates on average. Nevertheless, the mean residence time does not exceed the terminal time, i.e.~the settling time from a quiescent fluid, by a factor larger than four. An exception are simulations with only a slightly super-critical Rayleigh number ($\sim 10^4$), for which stationary convection develops and some particles become trapped indefinitely. 2D simulations of the same problem overestimate the flow-particle interaction -- and hence the residence time -- for both high and low Rayleigh numbers, which stresses the importance of using 3D geometries for simulating particle-laden flows. We outline how our model can be used to explain the depth changes of crystal size distribution in sedimentary layers of magmatic intrusions that are thought to have formed via settling of a crystal cargo, and discuss how the micro-structural observations of solidified intrusions can be used to infer the past convective velocity of magma.
\end{abstract}

\begin{highlights}
\item Large-scale circulation acts against particle settling for Rayleigh number below $10^7$.
\item For $Ra>10^7$, settling is described as a monotonic transition between two regime limits.
\item Crystal-size distribution in sediments may be indicative of the velocity of magma.
\item 2D simulations tend to overestimate the particle residence time.
\end{highlights}


\maketitle

\section{Introduction}

Convecting fluids often contain small particles. These can be ash particles or dust and pollutants in the atmosphere \citep{Schwaiger2012,Helbig2004} or in volcanic clouds \citep{Lemus2021}, micro-organisms and micro-plastics in lakes and oceans \citep{Ruiz2004, Malygina2021}, sand particles in an estuary \citep{Syvitski1985}, or crystals nucleated in solidifying liquids such as magma oceans \citep{ElkinsTanton2012,Solomatov2015} and magma chambers \citep{Holness2020}, or the Earth's outer core \citep{Koyaguchi1990}. In all the above situations, it is important to estimate the particle residence time, i.e.~the time particles spend within the host fluid before being deposited into a growing sedimentary layer at the bottom of the convecting layer (or at its top in the case of light crystals and bubbles of volatiles). While particle sinking has been extensively studied within the vast literature on particle-laden turbulent flows \citep{Wang1993,Tom2019,Pasquero2003}, the framework of thermally-driven flows is much less explored \citep[for experimental studies, see][]{Martin1989, Lavorel2009, Sturtz2021}. 

In first approximation, one could assume that the particle residence time $\tr$ is equal to $H/\avt$, where $H$ is the system height and $\vt$ is the particle Stokes velocity (also referred to as the terminal velocity). However, such assumption ignores any mutual interaction of particles, and also any possible effects the fluid flow can have on the settling behaviour. Slower or faster settling due to the presence of nearby particles \citep[respectively the hindering and collectivity effects, see][]{Culha2020} is important only when crystal concentration is large. The interplay between fluid flow and particle settling exists also in dilute suspensions, and is analyzed here.

Two limiting cases can be identified for the dynamics of inertial particles in dilute suspensions. Depending on the ratio of particle Stokes velocity to the mean flow velocity, $\avt/\vm$, particle trajectories may either resemble the trajectories of fluid tracers (advection-dominated regime), or be Stokes-like (settling-dominated regime). While fluid motion is strongly imprinted in the particle trajectories from the first group, in the second group the relatively fast-sinking particles ignore fluid motion altogether, i.e.~their settling statistics is the same as if the particles were placed into a quiescent fluid.

In a cooling body of magma, both these situations can take place, depending on the crystal radius, the density contrast with respect to the background fluid, and on the assumed flow velocity. For example, for a mineral phase 20\% denser than the parental magma, a 1 mm crystal will fall under the advection-dominated regime, while a 1 cm crystal will be closer to the settling-dominated regime in a convecting magma chamber \citep[][hereafter `our previous study']{Patocka2020}. In accordance with the terminology coined in our previous study, we refer to these two limits as the ``dust-like'' ($\avt/\vm \ll 1$) and ``stone-like'' ($\avt/\vm \gg 1$) regimes.

The particle mass conservation equation imposes that the rate of particles leaving the suspension, $\mathrm{d}N/\mathrm{d}t$, is equal to the particle flux at the boundaries. This is given by $A v_t c_\mathrm{wall}$, where $A$ is the area of the bottom boundary and $c_{\mathrm{wall}}$ is the mean particle number concentration near the bottom wall (or, in the case of buoyant particles, near the top wall). For well-mixed particles with relatively small Stokes velocity, the concentration $c_\mathrm{wall}$ can be estimated as the time-evolving volume averaged concentration $N(t) \,/\, (AH)$. For stone-like particles with relatively large Stokes velocity $c_\mathrm{wall}$ is determined by the initial concentration, which is equal to $N_0 / (AH)$ for an initially uniform distribution. The number of particles in suspension, $N(t)$, then either evolves as $N_0 \exp(-\vt t / H)$, or as $N_0 (1-\vt t/H)$, where $N_0$ is the initial number of particles \citep[for a detailed analysis of these two scenarios, see our previous study][]{Patocka2020}.

In both these limiting cases, the increment in the number of sedimented particles $\mathrm{d}N$ is proportional to $\vt \mathrm{d}t$, but the integral quantity $\tr$ is generally not equal to $H/\vt$. The average residence time $\tr$ depends on whether the linear or the exponential settling law applies. Distinguishing between the two can be of importance in solidifying liquids: for a given initial distribution of crystal sizes in the initial crystal load, the structure of sediment will be different if the dynamics of particles is advection-dominated (dust-like), or settling-dominated (stone-like). 

Understanding magmatic processes from igneous textures is a promising field that is gaining increasing attention \citep{Jerram2018}, partly because the paradigm of magma chambers has recently shifted \citep[for a review, see][]{Sparks2019}.
Convection of magma is sometimes taken into account (see e.g., \citep{Holness2017}, where the micro-structure of the Shiant Islands main sill is analyzed). However, to our knowledge, no study has focused on the different dynamic regimes arising in the presence of a population of polydisperse particles with widely-varying Stokes velocities, that are suspended in the same background flow. In this work, we outline how the differences between advection- and settling-dominated regimes can be used to make inferences about past convective vigour in solidified intrusions.

Another problem is that settling can be either enhanced or suppressed due to interactions with the fluid flow, especially when the investigated particle types lie in between the dust- and stone-like end-members. In turbulent flows, heavy particles are ejected away from flow vortices and thus preferentially sample only certain flow structures \citep{Eaton1994}, while light particles (and small bubbles) are attracted toward flow vortices \citep{Calzavarini2008} (hereafter referred to as ``preferential sampling''). A different interaction is described in our previous study, where we demonstrate how the presence of large-scale circulation can delay the average settling of both heavy and light particle types \citep[see the `slow belt' in Fig.~10 of][]{Patocka2020}. In a simplified sense, preferential sampling is related to local flow structures, while the latter interaction depends on the global flow structure.

Although the effect of large-scale circulation on particle settling was decreasing for Rayleigh numbers ($Ra$) greater than $10^9$ in the 2D simulations from our previous study, the settling dynamics in the limit of extremely high Rayleigh number convection remained unclear. This is because the amplitude of preferential sampling seemed to increase with increasing convective vigor: The focusing of light particles in flow vortices, inhibiting particle transport toward the top boundary was positively correlated with the Rayleigh number. Similarly, the average speed-up of heavy particles deposition due to ejection from vortices was easier to detect for the highest investigated $Ra$. As a result, already for $Ra \gtrsim 10^{8}$ the settling statistics of light and heavy particles mutually differed, suggesting that the mean residence time in highly vigorous convective flows would strongly depend on the density ratio $\rho_\mathrm{p}/\rho_\mathrm{f}$, i.e. the ratio of particle-to-fluid density. The resulting model for particle settling was thus relatively complex, with $\avt\,/\,\vm$ not being the only one important control parameter.

In this follow-up 3D numerical study, we argue instead that the settling behaviour of particles in convective flows at high $Ra$ is much more symmetric with respect to the sign of the Stokes velocity (i.e.~for heavy vs light particles), and allows for a simple description of the mean residence time. We demonstrate that 2D flows may artificially increase the interaction between large-scale circulation and particle dynamics, and, more importantly, that preferential sampling does not affect the mean residence time in the particle parameter space of interest, at least up to the highest $Ra$ that we simulated ($10^9$).

In low Rayleigh number convection, on the other hand, the coupling between flow and particle dynamics is stronger. For stationary flows, i.e.~for near-critical values of $Ra$, certain particle types may develop regular trajectories with infinite residence times. This phenomenon is related to the well known behaviour of particles in cellular flows \citep{Stommel1949,Maxey1987}, and we analyze to what value of $Ra$ it limits the applicability of our high-$Ra$ model.

While we present a general treatment suitable for a variety of natural and industrial systems, the paper is largely motivated by the problem of crystallization of liquid silicates. In particular, the model parameters are tailored for pools of cooling magma, spanning from global primordial magma oceans \citep{Tonks1993,Solomatov2015} to present-day magma chambers \citep{Sparks2019}. The residence time of crystals in solidifying magma oceans is of primary importance for the long-term thermochemical evolution of the interior of planets \citep{Tosi2020}, and similarly it determines the composition of the rock that forms upon the freezing of a magma chamber \citep{Martin1989,Koyaguchi1990,Holness2020}. We analyze not only the mean value of the residence time, but also its underlying probability distribution. It differs significantly from one end-member regime to the other, which has consequences for the solid-liquid phase separation and sediment structure in these systems.

In Section \ref{sec:Met}, we specify the governing equations and describe key control parameters. In Section \ref{sec:Res}, the residence time of particles is presented for a broad range of flow and particle parameters. Deviations from the general model are presented in Section \ref{sec:lowRa}. In Section \ref{sec:Mch}, we apply the results to a simplified system representing a generic magma reservoir and comment on possible steps towards building a self-consistent model of a crystallizing magma. Our conclusions are summarized in Section \ref{sec:Sum}.

\section{Method}\label{sec:Met}

As in our previous study \citep{Patocka2020}, we solve the Boussinesq equations in the following non-dimensional form:
\begin{eqnarray}\label{NVSnonD1}
    \partial_{t} \bm{u} + (\bm{u} {\cdot} \bm{\nabla})\bm{u} &=& -\bm{\nabla} p  + \sqrt{\frac{Pr}{Ra}}\, \bm{\nabla}^2 \bm{u} + \theta\, \bm{\hat{z}}\\
    \bm{\nabla} \cdot \bm{u} &=& 0\\
    \partial_{\tau} \theta + (\bm{u} {\cdot} \bm{\nabla})\,\theta &=& \sqrt{\frac{1}{Pr Ra}} \,\bm{\nabla}^2 \theta ,
    \label{NVSnonD3}
\end{eqnarray}
where $\bm{u}$ is the fluid velocity, $p$ is the dynamic pressure, $\theta$ is the temperature deviation with respect to a reference value ($T_0$), and $\bm{\hat{z}}$ is a unit vector in the vertical direction. The Rayleigh and Prandtl numbers determine the nature of the flow, and they depend on the fluid properties:
\begin{equation}
Ra = \frac{\alpha g \Delta T H^3 }{\nu \kappa}, \quad Pr =\frac{\nu}{\kappa},
\end{equation} 
where $\nu=\eta/\rho_0$ is the kinematic viscosity, $\eta$ is the dynamic viscosity, $\rho_0$ is the mean mass density at the reference temperature $T_0$, $\alpha$ is the volumetric thermal expansion coefficient, $\bm{g}$ is the gravitational acceleration, $\Delta T$ is the temperature scale, $H$ is the thickness of the convecting layer, and $\kappa$ is the thermal diffusivity.

We systematically investigate basally heated convection in a statistically steady state for $Ra=[10^4 - 10^9]$ and $Pr=[1,10,100]$ (see Section \ref{sec:Mch} for a discussion of the expected Rayleigh numbers of freezing bodies of magma). The aspect ratio is 2 in both the x- and y-directions. Side-walls are no-slip boundaries with the exception of the x-direction in which the walls are open. The top and bottom boundaries are isothermal, with a constant temperature difference driving thermal convection. The resolution goes up to $1024 \times 1024 \times 512$ for $Ra=10^8$, and due to limited CPU resources we perform the $Ra=10^9$ simulations in a limited aspect ratio in the y-direction (resolution $2048 \times 512 \times 1024$). We choose basally heated convection because it is the most typical setup, well suited to become a reference point for future work. However, for the intermediate Prandtl number we also investigate statistically steady flows with internal instead of basal heating, because these better represent the temperature profile that develops during the transient cooling of magmatic reservoirs with a layer of insulating sediment at the bottom (see Appendix \ref{sec:IH}). 

The fluid carries inertial particles, whose trajectories are governed by pressure and friction forces from the surrounding fluid in combination with particle buoyancy. Under idealized conditions of spherically-shaped particles with small Reynolds number, the Lagrangian equation of motion for a massive particle reads \citep[for more details, see][]{Mathai2016, Patocka2020}:
\begin{align}\label{eqnonDMR}
\der{\bm{v}}{t}=\beta \frac{D\bm{u}}{Dt} + \frac{1}{\St}(\bm{u}-\bm{v}) + \Lambda \,\bm{\hat{z}}, 
\end{align}
where $\bm{v}$ is the particle velocity and the first term on the RHS contains the material derivative of the fluid velocity. The modified density ratio $\beta = 3\rho_\mathrm{f} /(\rho_\mathrm{f} + 2\rho_\mathrm{p})$ involves the density of the fluid $\rho_\mathrm{f}$ and the particle density $\rho_\mathrm{p}$, and stems from the added mass force \citep{Auton1987}. 

The Stokes number $\St$ and Lambda parameter $\Lambda$ are defined as follows:
\begin{equation}\label{eqStLBeta}
\St = \frac{\rc^2 \sqrt{\alpha g \Delta  T}}{3 \nu \beta \sqrt{H}}, \quad 
\Lambda = \frac{\beta{-}1}{\alpha \Delta T}.  
\end{equation}
The Stokes number describes the viscous friction acting on each particle due to its relative motion with respect to the fluid. It is a non-dimensionalization of the viscous response time $\tau_D = \rc^2 / (3 \nu \beta)$ that characterizes how long it takes for viscous friction to adjust the particle velocity to that of the fluid -- with respect to a typical flow time scale, here chosen to be the fluid free fall-time $\sqrt{H/(\alpha g \Delta T)}$. The parameter $\Lambda$ (hereafter buoyancy ratio) expresses the relative importance of particle buoyancy with respect to the thermally-induced buoyancy of the fluid. 

As discussed in Section \ref{sec:drift}, the particle response time of crystals in magma bodies is relatively small, and Eq.~\eqref{eqnonDMR} can be replaced with its first order Taylor expansion \citep{Maxey1987,Balkovsky2001}.
\begin{equation}\label{eqpert}
\bm{v} = \bm{u} - St\,\Lambda\hat{\bm{z}} + St\, (1-\beta) \frac{D \bm{u}}{Dt},
\end{equation}
where $St \Lambda$ is the non-dimensional Stokes velocity $v_\mathrm{t}$ (defined positive for sinking and negative for rising).

We uniformly distribute $10^7$ particles of 201 different types into a fully developed, three-dimensional, statistically-steady thermal convection, with each particle type represented by three values: $St$, $\Lambda$, and $\beta$. The initial velocity of all particles is set to the local velocity of the fluid. Two hundred different types of particles are obtained by evenly sampling $\rho_\mathrm{f}/\rho_\mathrm{p}$ and $\rc^2$; one particle type is reserved for fluid tracers.

Using Eq.~\eqref{eqpert} instead of \eqref{eqnonDMR} allows us to model particles that have density contrasts and sizes corresponding to crystals in a primordial, global magma ocean \citep{Solomatov2015}. In Table \ref{parpam}, the model parameters that are used to evaluate the particle control numbers $St, \Lambda$, and $\beta$ are summarized. Note that the model domain depth $H$ and crystal radius $\rc$ enter the Stokes number $St$ as $\rc^2 / \sqrt{H}$, and so e.g.~a 16 times smaller body of magma with crystals of half the original size would be represented by the same $St$. For further guidance on how to apply our results to various natural systems, see Section \ref{sec:Mch}.

\begin{table}[width=1.\linewidth,cols=4,pos=h]
\begin{tabular*}{\tblwidth}{@{} LLLL@{} }
\midrule
 Parameter & Symbol &  Value & Units \\
\hline
  Mantle depth  & $H$ & $2890$ & km \\
  Grav.~acceleration & $g$ & $9.8$ & m/s$^2$ \\
  Thermal expansivity$^a$ & $\alpha$ & $5{\cdot}10^{-5}$  & K$^{-1}$\\
  Thermal diffusivity$^b$ & $\kappa$ & $5{\cdot}10^{-7}$& m$^2$/s  \\
  Kinematic viscosity$^c$ & $\nu$ & $[1,10,100]\cdot\kappa$ & m$^2$/s \\
  Temperature contrast$^d$ & $\Delta T$ & 1 & K \\
  Crystal size$^a$ & $\rc$ & $0.5 - 10$ & mm \\
  Density ratio & $\rho_\mathrm{p}/\rho_\mathrm{f}$ & $0 - 2$ & -- \\
\hline
\bottomrule
\end{tabular*}
\caption{Model parameters used for evaluating $St, \Lambda$, and $\beta$. $^a$ From \citet{Solomatov2015}; $^b$ from \citet{Huaiwei2015}; $^c$ see \citet{Karki2010} for typical viscosities of silicate liquids at high pressure and temperature; $^d$ see e.g. \citet{Lebrun2013} and \citet{Nikolaou2019} for typical temperature contrasts during the evolution of magma oceans.}\label{parpam} 
\end{table}

The above described model system is numerically simulated by means of the Eulerian-Lagrangian code \textit{ch4-project} \citep{Calzavarini2019}. The model setup of this paper differs from our previous study in three aspects: i) thanks to Eq.~\eqref{eqpert} we can now directly cover the desired region in the particle parameter space, see Fig.~\ref{fig:StL} in Appendix \ref{sec:drift}, ii) three-dimensional simulations are performed instead of two dimensional ones, and iii) we include also fluids with a small convective vigor. In particular, we include simulations with only a slightly super-critical $Ra$.

\section{Residence Times}\label{sec:Res}

\begin{figure}
    \centering
    \includegraphics[width=0.48\textwidth]{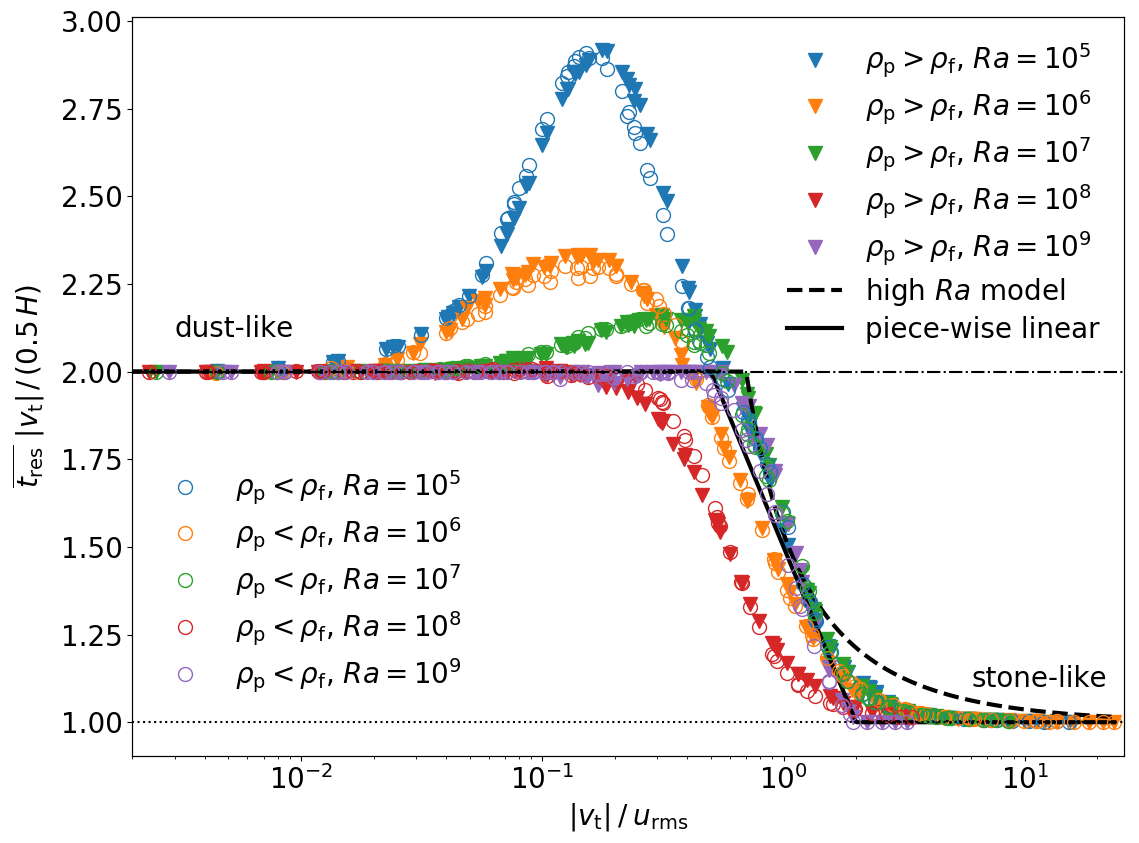}
    \caption{Mean residence time of the particles, normalized by the mean terminal time $0.5\,H/\avt$. Triangles represent heavy particles, i.e.~the particle types with $\rho_\mathrm{p}/\rho_\mathrm{f} \in (1{-}2]$, while open circles stand for light particles, $\rho_\mathrm{p}/\rho_\mathrm{f} \in [0{-}1)$. Black line represents the high-$Ra$ model towards which the results converge when $Ra$ is increased.}
    \label{fig:ResT}
\end{figure}

In Fig.~\ref{fig:ResT}, we plot $\mrt \,/\, (0.5 \, H / \avt)$, which is the mean residence time, $\mrt = \int_0^{\infty} N(t) \,\mathrm{d}t / N_0$, normalized by the mean residence time that would be obtained for a quiescent fluid (averages are performed over a given particle type -- the Stokes velocity is the same for all particles from each considered ensemble). In other words, $\mrt / (0.5 \, H / \avt)$ quantifies the importance of the fluid flow on the particle settling behaviour.

\begin{figure}
    \centering
    \includegraphics[width=0.48\textwidth]{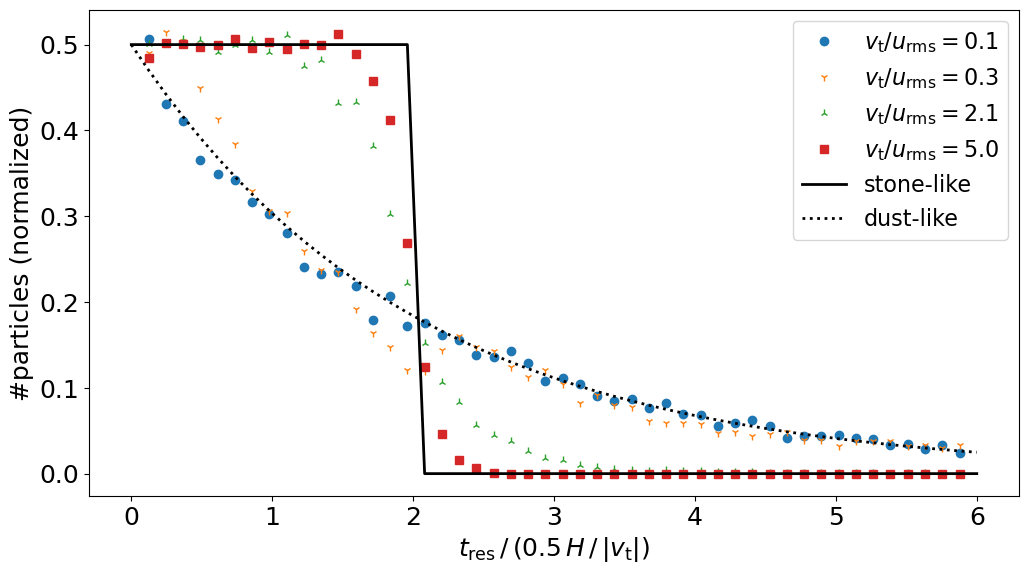}
    \caption{Probability density function of $\tr$ for $Ra=10^6$ and $Pr=1$. The analytical distributions for the dust-like (dotted black line) and stone-like (solid black line) regimes stem from Eq.~\eqref{eqmasscon}, upon considering the concentration at the settling front to be either $N(t)/H$ (dust-like regime) or $N_0/H$ (stone-like regime).}
    \label{fig:DisT}
\end{figure}

The two limiting cases of $\avt / \vm \gg 1$ and $\avt / \vm \ll 1$ are to be understood as follows: 

When $\avt / \vm \gg 1$, the particles do not effectively see the flow, because the fluid barely moves during the time it takes for these particle to reach the bottom (or top) of the container. The mean residence time $\mrt$ is equal to $0.5 H/\avt$ in this limit, because the particles are initially distributed uniformly throughout the container, and their mean height is thus $0.5 H$ at the time $t=0$ (see also the solid black line in Fig.~\ref{fig:DisT}). Note that $H$ is equal to 1 in the non-dimensional formalism, but we keep it in all our expressions in order to avoid confusions when using the resulting relationships in their dimensional form.

When $\avt / \vm \ll 1$, on the other hand, the particle dynamics are in the regime of a well-mixed suspension \citep{Martin1989}. These (dust-like) particles are perfectly stirred within the convective bulk, and effectively sink only when they are randomly transported to the thin boundary layer near the bottom (or the top, when $\rho_p/\rho_f < 1$). The initial positions of particles do not matter in this regime, because redistribution (mixing) within the model domain takes place in approximately one large-eddy turnover time, an interval much shorter than the terminal time of these particles. In Fig.~\ref{fig:DisT}, we show that the particle residence time in the dust-like regime follows an exponential distribution, with the probability density function being $\sim \exp (-\tr H /\avt)$, yielding $\mrt = H/\avt$ after averaging over the ensemble.

In between, for $0.02 \lesssim \avt/\vm \lesssim 2.0$, the large-scale circulation of the flow may alter the obtained particle dynamics, increasing the average residence time $\tr$. Simply put, this increase is caused by the locking of particles inside relatively stable convection rolls, as is described in \citet{Patocka2020} in more detail. The first important result in Fig.~\ref{fig:ResT} is that $\mrt$ is less affected by the large-scale circulation when $Ra$ increases, i.e.~that the ``slow belt'' we identified in our previous study disappears in the high-$Ra$ limit. 

The second important result is that there is no splitting of $\mrt$ for the heavy and light particles, not even when $Ra=10^9$ (circles vs triangles in Fig.~\ref{fig:ResT}). This is in disagreement with our previous 2D simulations (see e.g.~the blue vs yellow symbols in Fig.~\ref{fig:drifteq}, or Table II in our previous study). The reason of the disagreement, however, is not related to 2D vs 3D geometry (see Appendix \ref{sec:2D}). While the slow belt amplitude depends only on $Ra, Pr,$ and $\avt/\vm$, and moving within the $St,\Lambda$ parameter space along the $\avt$ isolines thus does not alter the way in which it affects $\mrt$, the strength of preferential sampling depends also on the value of $St$ itself, and can thus be different for two particle types with the same $\avt$ (see also Eq.~\ref{eqdivv2}). For this reason, preferential sampling was important in our previous study, but it plays a negligible role in the present simulations, in which $St$ and $\Lambda$ are chosen to match the values of interest in nature, and for which the range of $St$ is smaller by several orders of magnitude when compared to our previous study (cf.~Fig.~\ref{fig:StL} in Appendix \ref{sec:drift}).

The independence of $\mrt$ on the sign and value of $1-\beta$, cf.~circles vs triangles in Fig.~\ref{fig:ResT}, directly indicates the unimportance of preferential sampling, and thus also of the term $St\, (1-\beta) {D \bm{u}}/{Dt}$ in Eq.~\eqref{eqpert} (see also the expression for $\nabla \cdot \bm{v}$ and the related discussion in Appendix \ref{sec:drift}). Indeed, we have repeated one $Ra=10^4$ and one $Ra=10^8$ simulation, replacing Eq.~\eqref{eqpert} with even simpler particle dynamics in the form of $\bm{v} = \bm{u} - St\,\Lambda\hat{\bm{z}}$, and the settling statistics were comparable.

Below we construct an idealized model that roughly captures the observed residence times. The particle mass conservation imposes that the rate of particles leaving the suspension is
\begin{equation}\label{eqmasscon}
    \der{}{t} N(t) = -\avt \,A\, c_\mathrm{wall},
\end{equation}
where $c_{\mathrm{wall}}=\langle c(z {\to} 0,t) \rangle$ is the mean particle concentration at the wall, for which we use the following estimation:
\begin{equation}\label{eqcwall}
    c_\mathrm{wall} = \frac{N(t)}{A\, (H-t\, \vsh)}.
\end{equation}
The shrinking velocity $\vsh$ describes how the particle cloud, i.e.~the region that contains the $N(t)$ particles, changes in volume with time. Its value is different for each particle type, approaching 0 in the dust-like limit (particles always occupy the entire volume $A H$), and $\avt$ in the stone-like limit (the volume $A \vt t$ at the top of the tank is particle-free). With the help of Eq.~\eqref{eqcwall}, the solution of Eq.~\eqref{eqmasscon} can be obtained:
\begin{equation}\label{eqmodsol}
    N(t) = N_0\left(1 - \frac{t\, \vsh }{H}\right)^{\frac{\avt}{\vsh}},
\end{equation}
which gives the desired $N_0 \exp(-\vt t / H)$ and $N_0 (1-\vt\,t/H)$ respectively in the limits of $\vsh \to 0$ and $\vsh \to \avt$.

The normalized residence time is then obtained by integrating the solution \eqref{eqmodsol},
\begin{equation}\label{eqsoltres}
\frac{\avt}{0.5\,H}\,\mrt = \frac{2\avt}{\vsh+\avt}.
\end{equation}
Finally, we assume that the shrinking velocity can be approximated as
\begin{equation}\label{eqmodel}
\vsh = \left\{ \begin{array}{ll} 
\avt - \gamma \,\vm, & \mathrm{if}\; \avt > \gamma\, \vm \\[0.3cm]
0, & \mathrm{if}\; \avt \leq \gamma\,\vm 
\end{array} \right. 
\end{equation}
where $\gamma$ is a constant representing the mixing of particles - it mimics the fact that some particles are transported above the settling front due to the presence of fluid flow. The second branch in Eq.~\eqref{eqmodel} expresses the constraint $\vsh>0$, because the model domain does not expand in response to mixing. The normalized residence time is then
\begin{equation}\label{eqfinal}
\frac{\avt}{0.5\,H}\,\mrt = \left\{ \begin{array}{ll} 
\frac{2}{2-\gamma \,{\vm}\,/\,{\avt}}, & \mathrm{if}\; \avt/\vm > \gamma \\[0.3cm]
2, & \mathrm{if}\; \avt/\vm \leq \gamma 
\end{array} \right. 
\end{equation}
Based on the $Ra=10^9$ simulations, the best-fit choice for $\gamma$ is $0.7$. The corresponding solution is plotted with black dashed line in Fig.~\ref{fig:ResT}.

As an alternative to the high-$Ra$ model presented above, we suggest also an empirical law. The black solid line in Fig~\ref{fig:ResT} shows a piece-wise linear model that goes from 2 to 1, with $\avt/\vm=0.5$ and $\avt/\vm=2.0$ marking the transition on the $x$-axis. The empirical law provides a slightly better fit and is easy to remember, but lacks any physical insight.

Note that the probability density distributions of $\tr$ show certain deviations from the idealized end-members for values as small as $\avt/\vm=0.1$ and as high as $\avt/\vm=5.0$   (cf.~the black solid and dotted lines in Fig.~\ref{fig:DisT}).

The above model is derived for an initially uniform distribution of particles. Care must be taken when the model is applied to different settings. For instance, let us consider a case in which all heavy particles are near the top boundary at the time $t=0$ (and all light particles are near the bottom boundary). For such a setup, the normalizing factor becomes $H \,/\, \avt$ instead of $0.5\,H \,/\, \avt$, because the average distance from the boundary is now $H$ instead of $0.5\,H$. 

The end-member values of the normalized mean residence time in this modified setup are 1 for the dust-like limit and 1 for the stone-like limit, i.e.~$\mrt := H / \avt$ regardless of the value of $\avt/\vm$. The change of the normalized value from 2 to 1 in the dust-like limit is caused by the fact that the starting positions of particles are nearly irrelevant in this regime: their mixing time scale is negligible compared to their mean residence time. Therefore, $\mrt$ of the dust-like particles is only little sensitive to the initial positions, while the normalizing factor changes by a factor 2 (the mean terminal time is $0.5\,H \,/\, \avt$ for the uniform distribution, and $H \,/\, \avt$ for emplacement at the roof). In this study we report results for the initially uniform distribution, because then $\mrt$ of the dust-like particles is not at all affected by the time it takes to mix the particles throughout the model domain. For uniformly distributed particles, deviations of $\mrt$ from $(H \,/\, \avt)$ are strictly associated to flow-particle interactions (see Section \ref{sec:lowRa}). For the stone-like limit, on the other hand, the mean residence time $\mrt$ is bonded to the initial positions of particles, making the normalized value of 1 independent on the initial set-up. Note, however, that the distribution of $\tr$ among the particle population is always different in the dust-like and stone-like regimes: the distribution for the dust-like particles will be the one plotted in Fig.~\ref{fig:DisT} regardless of the initial setup, while for stone-like particles emplaced at the roof the distribution will change, resembling a delta function.

Note also that $\mrt$ is the mean residence time, and not the time it takes for all the particles to leave the suspension. In certain applications, the latter may be of interest \citep[e.g.][]{Holness2017}. In the dust-like regime, even after five terminal times the fluid still carries a non-negligible fraction of particles, while in the stone-like regime the fluid becomes particle-free shortly after reaching $t = H \,/\, \avt$ (Fig.~\ref{fig:DisT}).

\section{Low Rayleigh number convection}\label{sec:lowRa}

The above description of particle settling is limited to convection with $Ra \gtrsim 10^8$. For lower convective vigor, large-scale circulation may prolong the mean residence time (slow belt). Because the convective vigour of cooling magmatic bodies is unknown (see Section \ref{sec:Mch}), we briefly discuss also the low-$Ra$ simulations. 

In Table \ref{tabmax}, the maximum value of $\mrt \avt \,/\, (0.5\,H)$ is provided for the different flow parameters ($Ra = [10^4 - 10^9]$, $Pr = [1, 10, 100]$). There is a convergence toward the high-$Ra$ model regardless of the value of $Pr$, but the slow belt amplitude is different for different $Pr$, because large-scale circulation and the thickness of plumes in particular both depend on the Prandtl number $Pr$.

\begin{table}[width=1.0\linewidth,cols=4,pos=h]
\begin{tabular*}{\tblwidth}{@{} LLLL@{} }
\midrule
 Ra  & $Pr=1$ & $Pr=10$ & $Pr=100$ \\
\hline
  0  & \multicolumn{3}{c}{quiescent fluid: 1.0} \\
  $10^4$ & >7.1 (98\%) & >7.9 (93\%) & >16.1 (79\%) \\
  $10^5$ & 2.3 & 2.9 & 3.8 \\
  $10^6$ & 2.1 & 2.3  & 3.3\\
  $10^7$ & 2.1 & 2.2 & 2.5  \\
  $10^8$ & 2.1 & 2.0 & 2.3 \\
  $10^9$ & 2.1 & 2.0 &  \\
  $\infty$ & \multicolumn{3}{c}{high-$Ra$ model: 2} \\
\hline
\bottomrule
\end{tabular*}
\caption{Maximum of the normalized mean residence time.
 For $Ra=10^4$, not all the particles have settled by the end of the simulation -- the provided value of $\mrt$ is only a lower bound. The number in parenthesis shows the percentage particles that have settled, the settling rate of the remaining ones is nearly zero because they are trapped.}\label{tabmax}
\end{table}

\begin{figure*}[t]
    \centering
    \includegraphics[width=0.9\textwidth]{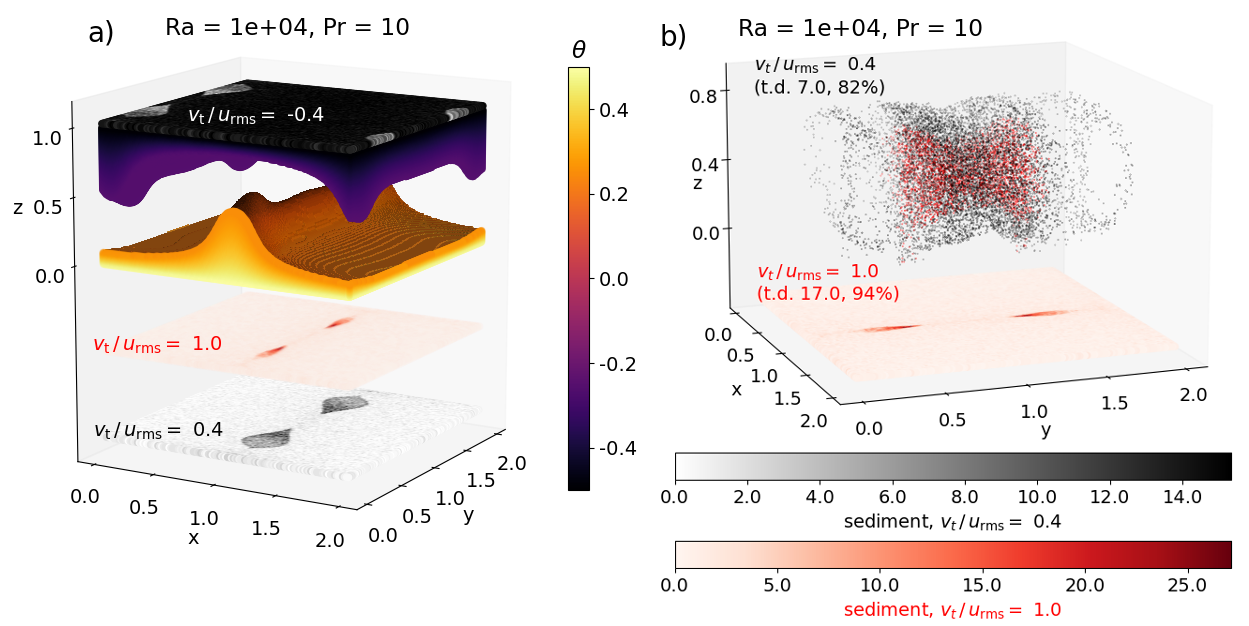}
    \includegraphics[width=0.9\textwidth]{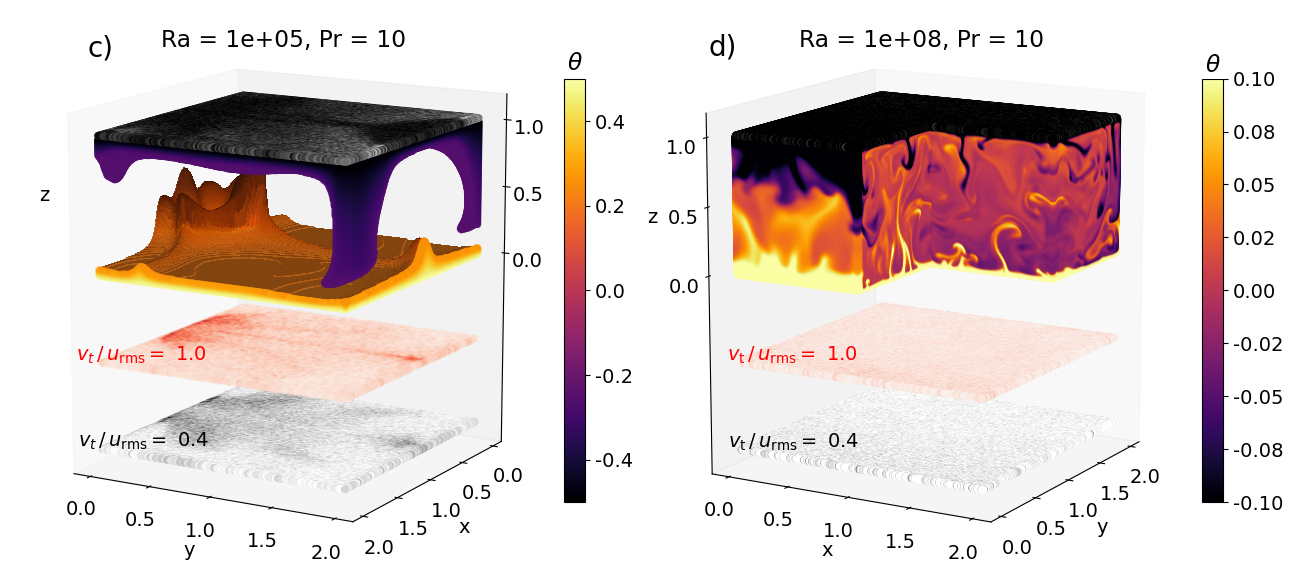}
    \caption{Temperature field and the spatial distribution of settled particles for selected $\vt/\vm$ ratios. a) Stationary convection with $Ra=10^4, Pr=10$. The non-dimensional temperature field is thresholded at 0.4 to show the upwellings, and at -0.4 to show the downwellings. The isosurface $T=0.4$ is shaded for better visibility. The 2D projections show the normalized distribution of settling events for $\vt/\vm = 0.4$ (black and white, floor of the plot), $\vt/\vm = -0.4$ (roof of the plot, inverse black and white), and for $\vt/\vm = 1.0$ (red). The color scales are shown below the plot in panel b). b) Particles that remain in suspension at the time $t=300$ for two heavy particle types: $\vt/\vm = 0.4$ (black) and $\vt/\vm = 1.0$ (red). Terminal distance and percentage of settled particles are provided in parentheses for each type. For a comparison of the viewpoint angle, the projection of settling events for $\vt/\vm = 1.0$ is repeated. c), d) Similarly as in panel a), only the Rayleigh number is increased. For $Ra=10^8$, we show a cut through the entire temperature field, with the color scale being clipped for better visibility.}
    \label{fig:c3d}
\end{figure*}

In Fig.~\ref{fig:c3d}a, we show the nearly stationary flow that is obtained for $Ra=10^4$ and $Pr=10$, i.e.~for only a slightly supercritical Rayleigh number. As a first approximation, the flow has a 2D structure -- it is dominated by two convection rolls that are aligned parallel to the $y$-axis, separated by a central upwelling structure (see the shaded ``ridge'' formed by the temperature isosurface that is parallel to the $y$-axis, with $x$-coordinate equal to 1).

The percentage of particles that become suspended inside the convection rolls decreases with increasing $\avt/\vm$, and the respective clusters of particles are increasingly closer to the central ridge (see Fig.~\ref{fig:c2d} in Appendix \ref{sec:2D} for the clustering of particles in 2D flows). Such spatial distribution is similar to what was reported already in the work of \citet{Stommel1949}, and analyzed later in more detail by \citet{Maxey1987} \citep[see also the ``retention zone'' in][]{Weinstein1988}. As a consequence of this behaviour, the percentage of particles whose settling is impeded by the presence of convection decreases with increasing $\avt/\vm$, but the normalized residence time of particles trapped inside the rolls increases with $\avt/\vm$ (the residence time of particles in the retention zone is similar across different particle types, but the normalization factor differs). 

Note that, although the concentration of particles is non-uniform, it is not locally increased with respect to the initial concentration. The non-uniformity is caused by the fact that particles are prevented from settling in some regions of the flow, while in neighbouring regions they are not. It is a different mechanism from preferential sampling, in which light particles from a certain region become focused near a flow vortex, forming a localized cluster that generates a sharp peak in the concentration field \citep{Patocka2020}.

A closer look at Fig.~\ref{fig:c3d}a reveals deviations from the 2D symmetry, and these are important for understanding the detailed picture. The central upwelling is stronger near y=0 and y=2.0, as illustrated by the elevated height of the shaded temperature isosurface. Particles above and in between the two peaks reside in the fluid the longest (Fig.~\ref{fig:c3d}b), their residence time is approaching infinity. Note that obtaining $\mrt \to \infty$ does not imply that all the particles of the given type would stay indefinitely in suspension -- a significant fraction of particles settle from the convective bulk for all the considered particle types. It is only the (small) fraction that are trapped inside the flow that makes the average value of $\tr$ go to infinity.

Most of the settling events take place below the ridge, particularly in the region where the flow gradient along y-direction is the strongest. The concentration of settled particles in these spots is up to 26 times higher than average value for particle type with $\vt/\vm=1.0$, and up to 15 times higher for particles with $\vt/\vm=0.4$ (respectively the red and black and white planes in Fig.~\ref{fig:c3d}a).

In Fig.~\ref{fig:c3d}c, we show the statistically steady flow with $Ra=10^5$ and $Pr=10$. The settling behaviour for $10^4 \lesssim Ra \lesssim 10^7$ can be described in a similar manner as the slow belt in our previous 2D study. The main mechanism acting here to retard the settling of particles is the uplifting force of plumes that form within a stable cluster, i.e.~whose origin is close to one localized region. Therefore, the mechanism described in \citet{Patocka2020} does exist also in 3D geometry, but is limited to lower Rayleigh numbers when compared to 2D simulations.

The concentration of settling events is increased below clusters of upwellings for heavy particles, and inversely for light particles (see the distributions of settling events in Fig.~\ref{fig:c3d}). This counter-intuitive result, reported already in our previous study and independently also in the 2D study of \citet{Xu2020}, thus remains valid in 3D geometry. Although plumes themselves act against the sinking motion of particles, the region from which they originate acts as a major particle sink -- the fluid flow near the floor is dominantly oriented toward the base of dominant plume clusters, and most of the solid material is thus deposited there. This is because fluctuations that encompass the birth of new plumes often cause particles to fall into the boundary layer. Only those particles that become entrained into a plume reside in the fluid anomalously long. The average residence time is then determined by the different likelihood of such interaction for the different particle types \citep[for a more detailed analysis, see][]{Patocka2020}.

For higher Rayleigh numbers, the flow-particle interactions disappear and the distribution of settling events becomes uniform (Fig.~\ref{fig:c3d}d). Note that for $Ra=10^9$ and $Pr=10$ the slow belt amplitude was the largest in our previous study (Table II, sets B therein), while in 3D it already decreases to zero for $Ra \approx 10^8$. As analyzed in Appendix \ref{sec:2D}, this is because 2D geometry artificially enhances the ability of convection rolls to prevent particles from reaching the boundary layers of the flow.

\section{Cooling Magma Reservoir}\label{sec:Mch}

The presented results are valid for spherical particles with time-constant density and radii, emplaced uniformly into a statistically steady flow. As such, the model setup is far from the natural systems of cooling magma bodies, in which the crystals spontaneously nucleate and grow in a transient flow, with the background fluid being confined in a shrinking domain. Nevertheless, solidifying magma is a thermally convecting suspension, and the dynamics that we observe are thus to some extent applicable also to these natural systems. Below we outline how to do so.

The primary control parameter in this study is the $\avt/\vm$ ratio. The Stokes velocity of natural crystals depends on the crystal shape, radius, density, and on the magma viscosity. While these parameters are usually known to a limited precision, their ranges are relatively well established and can be estimated for any system of interest. Note, for example, that the reduction of settling velocity caused by crystal shapes is typically below $10\%$ for crystal shape aspect ratios up to 4 \citep{Kerr1991}. In order to make inferences about crystal dynamics, the remaining and usually unknown parameter to estimate is the mean flow velocity $\vm$.

The Rayleigh number describing a reservoir filled with low-viscosity basaltic magma is typically huge, reaching $10^{29}$ for a global magma ocean \citep{Solomatov2015} and $10^{17}$ for large magma chambers \citep{Clark1987}. Perhaps the most uncertain parameter entering such estimates is the driving temperature contrast $\Delta T$. During the 90s, there was a sharp debate that boils down to whether $\Delta T$ becomes so small as to entirely stop convection in a cooling (freezing) magma chamber \citep{Marsh1989a,Huppert1991,Marsh1991}. The essential idea behind a small $\Delta T$ is that a highly viscous stagnant lid is continuously forming above the solidification front, reducing the temperature contrast driving convection in the liquid \citep[see Fig.~4 in][]{Brandeis1986}.

While being aware of the concluding remarks by \citet{Huppert1991}, who argue for the general likelihood of vigorous convection in the systems of interest, we point out one additional argument in favor of reducing the effective $\Delta T$. If the nucleated crystals grow most of their volume in the thermal boundary layer of the fluid, then the released latent heat should significantly reduce the temperature contrast that develops within the fluid. In fact, in the limit of the classical Stefan problem, the heat subtracted from the liquid is entirely compensated by the latent heat generated within the solidification front, i.e.~the temperature contrast driving convection is zero as the fluid stays at the liquidus at all times. Indeed, such formulation of the Stefan problem applies only to single-component systems, and assumes that all the newly forming solid material is attached to the top-down growing phase boundary -- far from the more complex case of magma chambers. Nevertheless, some $\Delta T$ and thus convective vigor reduction caused by the latent heat release should still be expected, and this point is not addressed in detail in the above mentioned debate.

Here we merely acknowledge that the convective vigor of a cooling magma reservoirs is, to a large extent, unknown \citep{Holness2020}. For this reason, we explore both vigorously and only moderately convecting fluids. Moreover, the paradigm of magma chambers has shifted in recent years, as there seem to be only little evidence for the existence of large volumes of liquid magma being enclosed in separate reservoirs within the crust \citep[for a review, see][]{Sparks2019}. Magmatic suspensions are thus likely to be encapsulated in smaller domains connected via mushy zones, which further complicates the estimates of the mean flow velocity $\vm$.

In this study we aim to provide a guidance as to what suspension dynamics to expect for a given $\vm$, and, conversely, how to infer $\vm$ from the petrological record in solidified intrusions.

\begin{figure}
    \centering
    \includegraphics[width=0.48\textwidth]{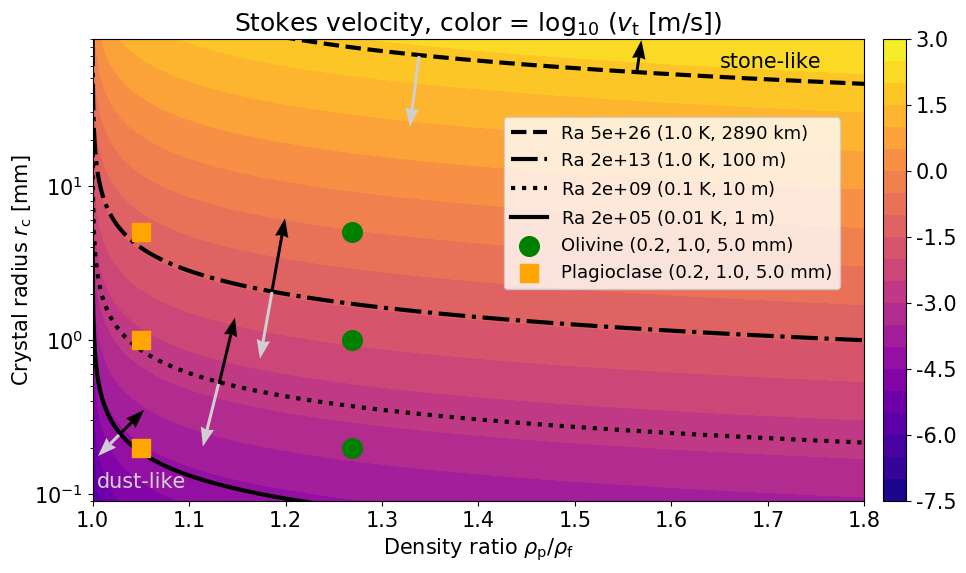}
    \caption{Stokes velocity of crystals in a convecting magma reservoir as a function of the density contrast $\rho_\mathrm{p}\,/\,\rho_\mathrm{f}$ and crystal radius $\rc$. Particle parameters other than explicitly specified are taken from Table \ref{parpam}, Prandtl number is 100 for all cases. Black lines mark the $\avt = \vm$ isolines for selected values of $\vm$, i.e.~these lines crudely correspond to the dust-stone transition for systems with different convective vigor, as indicated by the black (stone-like) and gray (dust-like) arrows.}
    \label{fig:Mch}
\end{figure}

In Fig.~\ref{fig:Mch}, we plot the Stokes velocity of particles for the typical density contrasts and the typical crystal radii characterizing magmatic systems. Black lines show the isolines corresponding to $\avt = \vm$ for selected $\vm$ values, that are obtained using a theoretical $\vm=\vm(Ra,Pr)$ scaling law \citep{Ahlers2009}, with several different values of $Ra=Ra(\Delta T, H)$. The choice of $\Delta T$ and $H$ spans the range from a vigorously convecting global magma ocean ($\Delta T=1$ K, $H=2890$ km) to only moderately convecting, small capsule of magma ($\Delta T=0.01$ K, $H=1$ m). As a crude approximation, the isoline $\avt = \vm$ can be taken as the division line between the dust-like and stone-like regime (cf.~Fig.~\ref{fig:ResT}). Fig.~\ref{fig:Mch} thus indicates what type of end-member dynamics one should expect for a given crystal type. Green circles in Fig.~\ref{fig:Mch} mark 0.1, 1.0, and 5.0 mm large crystals of olivine that is 700 kg/m$^3$ denser than the parental magma of density 2600 kg/m$^3$, squares represent plagioclase crystals with density 2730 kg/m$^3$.

The density of olivine and plagioclase used in Fig.~\ref{fig:Mch} comes from \citet{Holness2017}, who analyze the Shiant Islands main sill. They speculate that the first ca.~10 m of the picrodolerite/crinanite unit (PCU) formed as a result of settling of olivine and plagioclase crystals that were transported into the sill from the deeper crust (``crystal cargo''). They estimate the build-up of 10 m of sediment to take 22-48 weeks, while the solidification front would move only 2.6-3.9 m within that time, and conclude that ``initial settling of the crystal load would have occurred rapidly relative to the upwards movement of the solidification front, forming an essentially isothermal pile on the sill floor, with negligible associated in situ crystallisation''.

In the first ca.~10 m of the PCU, there is a gradual reduction in the proportion of large grains in the population upwards in the stratigraphy (fining-upwards), while the overlaying tens of metres are coarsening upwards, which is thought to be a result of the interplay between residence times and grain growth. It is debated whether the fining-upwards segment was formed via settling under the conditions of static magma (as suggested by an abrupt change of Cr-spinel concentration), or via settling from a convecting magma, for which \citet{Holness2017} apply the model of \citet{Martin1989}, i.e.~the dust-like regime, to the entire crystal population.

Understanding magmatic processes from igneous textures is a difficult task \citep{Jerram2018}. The micro-structure analysis of \citet{Holness2017} is a promising tool in this regard, and a similar method was used also in other studies \citep{Holness2006,Holness2020}. Below we perform an exercise following the Discussion in \citet{Holness2017}. 

Olivine crystals with initially Maxwell-type distribution (cf.~Fig 11 in \citet{Holness2017}) are suspended into magma of viscosity 2 Pa s and height 100 m. The mean and variance of the distribution are set respectively to $D_0=0.1$ mm and $n=2$. The initial volume fraction is set to 3.75\%, in a rough agreement with olivine mode being $\sim 50\%$ at the base of the PCU, and $\sim 25\%$ at the top of the fining-upwards sequence (we assume the crystal load of olivine to match in volume the olivine found in the 10 m tall sediment).

We use Eq.~\eqref{eqmodsol} to compute the settling of a polydisperse population of particles, testing three values of the background flow velocity $\vm$: 1) $\vm=0$ cm/s, i.e.~settling from a static magma, 2) $\vm=1$ cm/s, for which all the crystals fall into the dust-like regime. This value corresponds to temperature contrast driving convection to be $1$ K, resulting in $Ra\sim 10^{12}$, and 3) $\vm=10^{-4}$ cm/s. This value is arbitrarily chosen in order to split the initial crystal size distribution by the dust-stone transition, and corresponds to $Ra\sim 10^5$.

\begin{figure}
    \centering
    \includegraphics[width=0.48\textwidth]{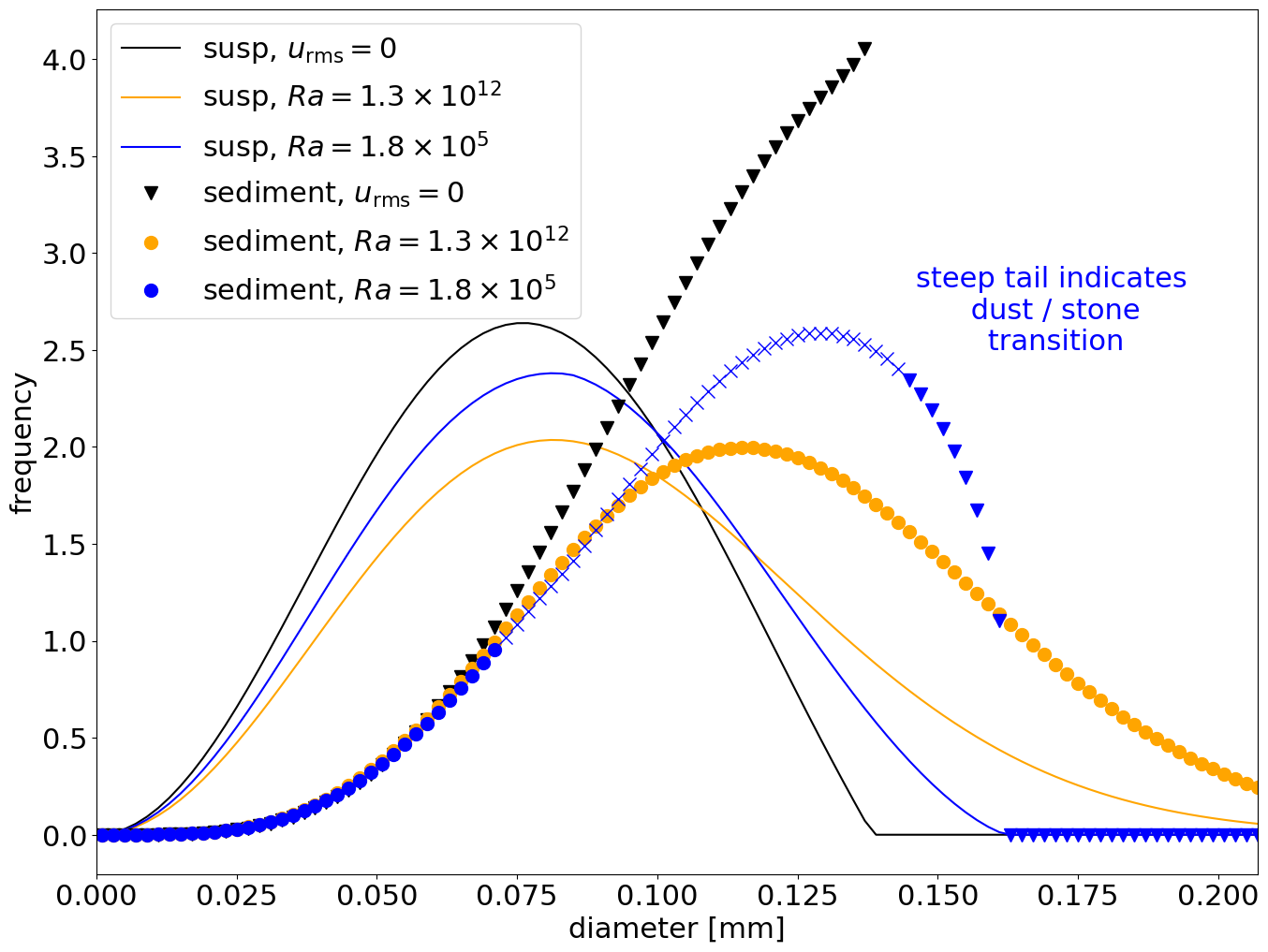}
    \caption{Crystal size distribution in suspension and sediment for a simple application that follows the Discussion in \citet{Holness2017}. We test three values of the fluid mean velocity $\vm$, corresponding to a static magma (black), vigorously convecting magma (orange), and to an intermediate case (blue). Solid lines show the size distribution of particles remaining in suspension at the time $t=40$ weeks, at which more than 50\% of all the particles have settled in all three cases. Symbols display the size distribution in the currently forming layer of sediment. Triangles represent particles with $\vt>2.0\, \vm$ (stone-like), circles show particles with $\vt<0.5\, \vm$ (dust-like), and crosses mark the transition.}
    \label{fig:disthree}
\end{figure}

In Fig.~\ref{fig:disthree}, we plot the distribution of crystals remaining in suspension after 40 weeks of evolution. At this time, $67\%$ of all particles are already deposited when $\vm=0$ cm/s, while for $\vm=1$ cm/s it is only $50\%$. Note that this difference may play a role when comparing the rate of sediment build-up with the speed of propagation of the solidification front -- while \citet{Holness2017} favour the scenario of settling from convection, they use the relation for build-up from static magma from \citet{Farr2017}.

While the size distributions of crystals that remain in suspension are similar, the size distributions of the currently forming sedimentary layers differ markedly. While stone-like settling (black triangles) results in a steep increase in frequency until the maximum crystal size is reached, and then the frequency drops to zero, dust-like settling (orange circles) produces a sedimentary layer with a smooth size distribution (Maxwell-type with $n=4$ in this case). The intermediate case (blue symbols) has a steep tail, and its peak corresponds to crystal radii for which $0.5\lesssim \avt/\vm\lesssim 2.0$, i.e.~it corresponds to the dust-stone transition. The significant (potentially observable) difference in sediment micro-structure is a general result that is obtained regardless of the details of the grainsize distribution in the initial crystal load, as long as the comparison is performed in later stages of settling (i.e.~when at least ca.~60\% of all particles have already settled).

Here we do not aim to systematically explore the parameter space of gravitational sorting, but we would like to point to the possibility to do so. If the grain-size distribution is available as a function of depth, one can use Eq.~\eqref{eqmodsol} to fit the observations, with $\vm$ being a free parameter. This should at least indicate either settling- or advection-dominated regime, but it could also identify the actual value of convective velocity $\vm$. The terminal velocity $\vt$ scales as $\rc^2$, and that grain size distribution typically has a large variance. Moreover, individual grains may form cohesive clusters (polycrystals), i.e.~particles with effective size of millimeters to centimeters (cf.~Fig 9 in \citet{Holness2017}), and $\vt$ may thus span orders of magnitude, making it more likely that $\vt$ crosses $\vm$ somewhere in the recorded distribution. Note that if the frequency distribution of polycrystals in the sediment was counted, its peak could indicate the dust-stone transition even if the background magma was convecting strongly (cf.~the dash-dotted black line in Fig.~\ref{fig:Mch}). The script used to produce the results plotted in Fig.~\ref{fig:disthree} is provided in the Supplementary material.

Note also that the presence of multiple phases can provide an additional constraint on the dynamic regime. For example, if we emplace both olivine and plagioclase crystals with the same size distribution into the sill, the mutual particle fluxes differ dramatically after 40 weeks, with the olivine to plagioclase ratio being 1.55 for $\vm=1$ cm/s, 0.75 for $\vm=0$ cm/s, and 1.12 for the intermediate case (due to the much smaller density contrast of plagioclase, most of its crystals fall under the dust-like regime for both the non-zero $\vm$ cases). At the later stages of settling, however, the fluxes are small, and again the build-up rate must be compared against the estimated advance of the solidification front in order to test whether the argument for negligible in-situ crystallization is still justifiable.

Above the fining-upwards segment of the PCU, tens of meters are found to be coarsening-upwards, which is explained as a result of grain growth and progressive clustering of particles that reside in suspension as solidification proceeds. While accounting for spontaneous nucleation and growth within direct numerical simulations is a challenging task, Eqs.~\eqref{eqmasscon}, \eqref{eqcwall}, and \eqref{eqmodel} provide a simple framework for constructing future parametrized models that would account for grain growth, and may serve as a quantitative tool for testing the geological interpretations of the micro-structure in solidified intrusions. 

Note that in the exercise above we use the high-$Ra$ model regardless of the actual value of $Ra$, i.e.~we do not account for the slow belt when $Ra$ is low. An intriguing implication of the results presented in Section \ref{sec:lowRa} is that, for a near-critical $Ra$, certain group of particles could be trapped in the bulk of a cooling magma body. When the flow gets disrupted, e.g.~by the arrival of new magma, such particles could suddenly become released, forming a distinct sedimentary layer. While discontinuities in layering are sometimes attributed to the onset of chamber-wide convection \citep{Holness2006}, our results suggest that they could also form in response to disruptions of an already developed, low-$Ra$ convection.

\section{Discussion}

\textbf{Extrapolation.} The possibility to apply our results also to systems with $Ra>10^9$ is based on the following two observations: i) The effect of global flow structures (large-scale circulation) on particle settling decreases with increasing $Ra$, and seems negligible already for $Ra \gtrsim 10^8$. This trend was observed also in our previous 2D study (with modifications discussed in Appendix \ref{sec:2D}), in which the highest reached $Ra$ was $10^{12}$. It can be expected that as $Ra$ increases, any large-scale flow structures that could potentially impede particle settling would be subject to strong fluctuations and disturbances caused by small vortices, making it easier for the particles to sink through any such structures. It is important to note, however, that extrapolation to high-$Ra$ flows is subject to an open debate, and in particular the dynamics of fluid boundary layers is uncertain in extreme regimes \citep[see e.g.][]{Ahlers2009}. Note also, that the slow belt amplitude as a function of the Rayleigh number is not monotonic in 2D (over the range of $Ra\in[10^4-10^{12}]$, see Table \ref{tab2d} in Appendix \ref{sec:2D} and Table II in our previous study). Although we see a decreasing trend toward the highest $Ra$ that we simulate, it cannot be ruled out a priori that large scale circulation does not become more important again as $Ra$ increases even further - global coherent flow structures are expected also in a fully developed turbulent regime \citep{Ahlers2009}, and for $Pr=1$ a small but yet observable peak remained in the $\mrt$ curve for all the simulated $Ra$ (Table \ref{tabmax}). ii) Preferential sampling due to the centrifugal effect is not observed in the present study, because the particle response time $St$ is extremely small for crystals appearing in natural magmas. Moreover, flow vortices in 3D flows dissolve relatively fast when they are not confined to a 2D plane, making it more difficult to attract and maintain particles from any significant region.

\textbf{Convective mode.} For reasons stated in Section \ref{sec:Met}, we assume basally-heated convection in our direct numerical simulations. However, after hot magma is emplaced into a cavity, it begins to cool from all sides. If the layer of sediment forms sufficiently fast, consistently with our results for most cases, then the floor becomes insulated from the host rock \citep{Jarvis1994}. The system is then effectively cooled from above and from the sides -- for a large aspect ratio domain it establishes a temperature profile that resembles that of internally heated fluid in a statistically steady state \citep[see e.g.][]{Sturtz2021}. In Appendix \ref{sec:IH}, we thus briefly investigate internally instead of basally heated convective flows. A systematic investigation of transient flows in which cooling from all sides would be considered is outside the scope of the present paper, and would require also a treatment of the solidification front growing from the bottom.

\textbf{Re-entrainment.} Particle re-entrainment from the sedimentary layer back into the convective flow \citep{Solomatov1993a, Solomatov1993b} is not considered here (see also a more recent study by \citet{Sturtz2021}, who investigate re-entrainment of both heavy and light particles in the context of a volumetrically heated fluid). As discussed in detail in our previous study, the workings of re-entrainment of natural crystals need to be revisited to account for the effects of compaction, chemical bonding, and for the non-sphericity of crystals shapes. 

\section{Summary}\label{sec:Sum}

The residence time of particles in a dilute convecting suspension may be affected by preferential sampling of local flow structures, as well as by flow-particle interaction due to large-scale circulation. We find preferential sampling to play a negligible role in a thermally-convecting magma. Large-scale circulation does retard the settling of particles with $0.02 \lesssim \avt/\vm \lesssim 2.0$, but its influence decreases with increasing convective vigor, being negligible already for $Ra \sim 10^8$. Together, this allows for a simple, monotonic description of the settling dynamics in a thermally convecting fluid with a high-$Ra$ (Eq.~\ref{eqmodsol}).

Igneous textures often allow for various interpretations of the magmatic processes that formed them. Our Eq.~\eqref{eqmodsol} can be used to study the settling of a population of polydisperse particles, transported into a magma chamber or sill from the deeper crust in the form of crystal cargo. The dust-like and stone-like dynamic regimes give markedly different crystal size distributions in the later stages of settling, which can be used to constrain the convective velocity of the host magma. To illustrate this, we analyze how olivine crystals fractionate under the condition of a static, vigorously convecting, and only moderately convecting magma. The particle size distribution in the sediment is noticeably different, with dominant frequencies indicating the mean fluid velocity in the intermediate case, i.e.~when differently sized crystals from the investigated population fall under different dynamic regimes.

\section*{Acknowledgements}
We warmly thank Lucie Taj\v{c}manov\'{a} for discussions. NT acknowledges support from the Helmholtz Association, Project No.~VH-NG-1017. VP acknowledges support from OP RDE project No.~CZ.02.2.69/0.0/0.0/18\_053/0016976, International mobility of research, technical and administrative staff at the Charles University. The computations were carried out on the DLR cluster CARA.

\printcredits
\appendix
\section{Particle Drift Equation}\label{sec:drift}

In dimensional form, the Lagrangian equation of motion for a small massive spherical particle is \citep[e.g.][]{Mathai2016}:
\begin{equation}\label{eqmaxeydim}
\frac{d\bm{v}}{dt}= \beta \frac{D\bm{u}}{Dt} + \frac{1}{\tau_D}(\bm{u}-\bm{v}) + (1 - \beta )\bm{g},
\end{equation}
where $\tau_D  = \rc^2 / (3 \nu \beta)$ is the particle response time and $\rc$ is the particle radius.

If equation \eqref{eqmaxeydim} is multiplied by $\tau_D$ and one takes the limit for $\tau_D \to 0$, the solution $\bm{v} = \bm{u}$ is readily obtained. This implies that for small values of $\tau_D$ the particle velocity $\bm{v}$ should be close to the fluid velocity $\bm{u}$.

In the limit of small but non-vanishing $\tau_D$, a perturbative solution of the above equation can be derived. In other words, Eq.~\eqref{eqmaxeydim} can be viewed as a class of differential equations, yielding for each $\tau_D$ a different solution $\bm{v} := \bm{v}(\tau_D)$ that can be expanded into its Taylor series around the point $\tau_D=0$. We therefore consider that the solution $\bm{v}$ will be of the following form:
\begin{equation}\label{eqansatz}
\bm{v} = \bm{u} + \tau_D \bm{v}_1 + \mathcal{O}(\tau_D^2).
\end{equation}
Upon substituting this ansatz into Eq.~\eqref{eqmaxeydim}, and taking into account the difference between the time derivative in the particle frame $d/dt$ and the convective derivative $D/Dt$ \citep{Mathai2016}, the first-order correction is obtained:
\begin{equation}\label{eqpertapp}
\bm{v} = \bm{u} + \tau_D (1 - \beta )\left( \bm{g} -  \frac{D \bm{u}}{Dt} \right) + \mathcal{O}(\tau_D^2).    
\end{equation}
Note that $\tau_D(1-\beta)\bm{g}$ is the particle Stokes velocity $\bm{v}_\mathrm{t}$. Therefore, Eq.~\eqref{eqpertapp} says that the particle has a certain drift with respect to the fluid velocity $\bm{u}$. This drift is given by the particle Stokes velocity, in which the gravitational acceleration $\bm{g}$ is corrected for the background fluid acceleration. In the non-dimensional form, i.e.~upon dividing by the characteristic velocity scale $u^*=\sqrt{\alpha g \Delta T H}$, Eq.~\eqref{eqpertapp} reads
\begin{equation}\label{eqpertap}
\bm{v} = \bm{u} + St\,\Lambda + St\, (1-\beta) \frac{D \bm{u}}{Dt} + \mathcal{O}(St^2),
\end{equation}
which is the formula used in this study (drift equation). This equation is valid in the limit $St\ll1$ or equivalently $v_t\ll\Lambda$ (because $v_t=St\Lambda$). Note that even in this approximation particles can accumulate in specific regions of the flow. This can be seen by taking the divergence of the fluid velocity
\begin{eqnarray}
{\bm \nabla} \cdot \bm{v} &=& St(1-\beta)\left(\frac{\omega^2}{4} - \mathcal{S}^2 \right) \label{eqdivv1}\\ 
&=& St(1-\beta)(-\nabla^2 p + \partial_z \theta), \label{eqdivv2}
\end{eqnarray}
with $\bm{\omega} = \bm{\nabla} \times \bm{u}$ and  $\mathcal{S} = (\bm{\nabla} \bm{u} + \bm{\nabla} \bm{u}^T)/2$ being respectively the fluid vorticity field and the rate of strain tensor. Eq.~\eqref{eqdivv1} implies that particles heavier than the fluid ($\beta<1)$ accumulate (have a negative divergence of the velocity field) in strain dominated regions. Alternatively, Eq.~\eqref{eqdivv2} tells that particles accumulate in pressure maxima (typically occurring outside vortices), although in thermally driven flows this effect has to be compared with the local intensity of temperature gradients (particles clustering is favoured by strong negative thermal gradients, a condition typically occurring in thermal boundary layers).

Without the $St\, (1-\beta) {D \bm{u}}/{Dt}$ term in Eq.~\eqref{eqpert}, the particle velocity divergence is zero and particles thus cannot accumulate in any flow regions. This, however, does not mean that the particle concentration must be uniform within the model domain at all times (see Appendix \ref{sec:2D}).

\begin{figure}
    \centering
    \includegraphics[width=0.48\textwidth]{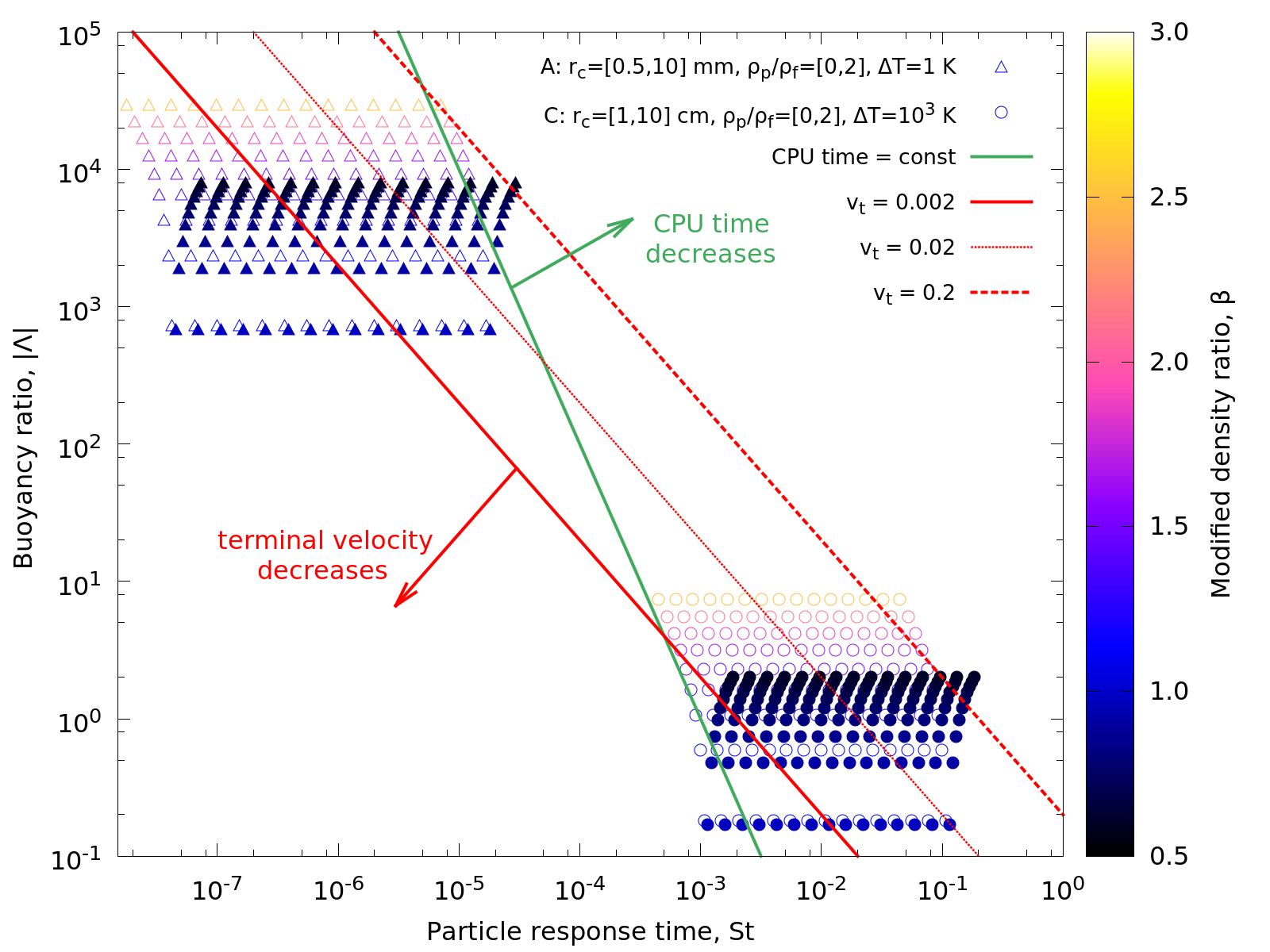}
    \caption{Each point represents a particle type in the $St, \Lambda, \beta$ parameter space. Filled symbols represent particle types that are denser than the fluid ($\beta<1.0$) and empty symbols represent light particles ($\beta>1.0$). Circles show simulation set C$^{10}_{50}$ that was used as the reference case in \citet{Patocka2020}. Triangles mark the region in the particle parameter space that is of our interest but was computationally unreachable with Eq.~\eqref{eqnonDMR}, computed in this study with the help of Eq.~\eqref{eqpert}. Red lines delineate the slow belt when $\vm=0.1$, i.e.~for $Ra=10^{10}$ and $Pr=50$.}
    \label{fig:StL}
\end{figure}

As explained in detail near Eq.~(13) in \citet{Patocka2020}, the numerical integration of Eq.~\eqref{eqmaxeydim} is subject to a constraint on the allowed time step $\Delta t$, such that $\Delta t \lesssim 0.1 \tau_D$ (or $\Delta t \lesssim 0.1 St$ in the non-dimensional formalism). This limits the computationally reachable area in the $\St, \Lambda$ space (green line in Fig.~\ref{fig:StL}). Using Eq.~\eqref{eqpertapp} instead of Eq.~\eqref{eqmaxeydim} relaxes the time step constraint.

\begin{figure}
    \centering
    \includegraphics[width=0.48\textwidth]{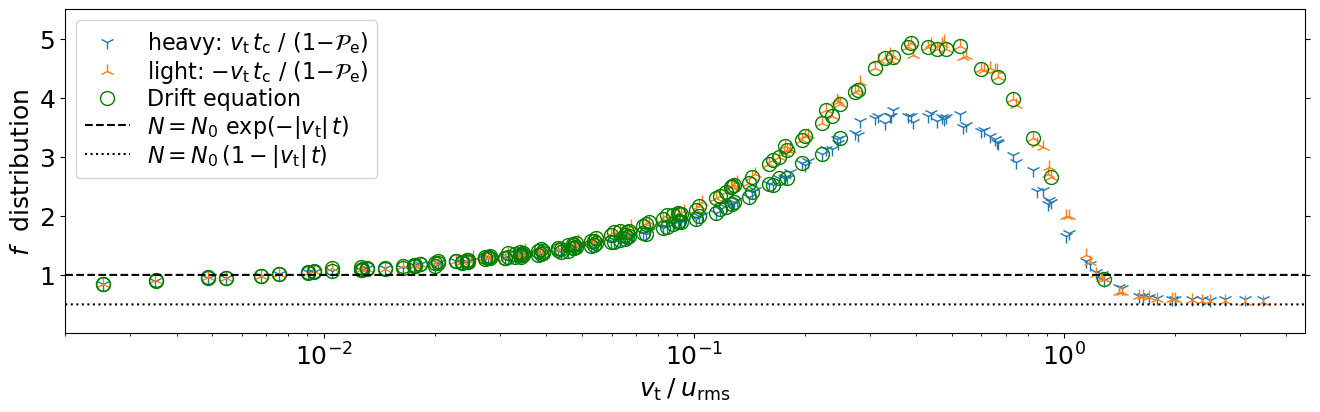}
    \caption{Average settling behaviour for the simulation set denoted as C$^{10}_{50}$ in Fig.~\ref{fig:StL}, represented by the $f$-distribution \citep[Eq.~19 in][]{Patocka2020}. Blue symbols stand for the particle types denser than the fluid (filled circles in Fig.~\ref{fig:StL}), orange symbols denote particles lighter than the fluid (empty circles in Fig.~\ref{fig:StL}). Particle types whose dynamics were computed using Eq.~\eqref{eqpert} instead of Eq.~\eqref{eqnonDMR} are circled in green.}
    \label{fig:drifteq}
\end{figure}

 In order to see whether the use of the drift equation affects particle settling behaviour, we recomputed the reference simulation set C$^{10}_{50}$ from \citet{Patocka2020}, using a larger numerical time step and employing Eq.~\eqref{eqmaxeydim} for particle types with $\tau_D > 10\, \Delta t$ (results for these particle types are circled in Fig.~\ref{fig:drifteq}). As is apparent from Fig.~\ref{fig:drifteq}, and from its comparison to Fig.~8 in \citet{Patocka2020}, the perturbative solution \eqref{eqpertapp} results in the same settling statistics as when Eq.~\eqref{eqmaxeydim} is employed, with the $f$-distribution being the same regardless of which equation was applied to advance the particle trajectories (the $f$-distribution has a similar meaning as half the normalized residence time plotted in Fig.~\ref{fig:ResT} in the main text).

Therefore, in this study we use the drift equation, Eq.~\eqref{eqpertapp}, that allows us to move to the desired region in the model parameter space, i.e.~to the region of interest for investigating cooling magma reservoirs (set A in Fig.~\ref{fig:StL}).

\section{2D Artefacts}\label{sec:2D}

Table~\ref{tab2d} is analogous to Table \ref{tabmax}, only this time the simulations are performed in 2D geometry, i.e.~it is assumed that the flow does not change along the y-direction. For identical flow parameters, the mean residence times are significantly higher, with the exception of the case with $Ra=10^7, Pr=100$. Note that the values of $\mrt$ are subject to uncertainties related to flow fluctuations. For the same $Ra$ and $Pr$, two thermal convection simulations may develop different flow structures and also the mean properties of the flow may slightly vary, e.g.~due to changes in the number or relative sizes of convection cells \citep{Patocka2018}. A thorough evaluation of these uncertainties is computationally expensive as it implies ensemble averaging over different simulations and it goes beyond our current numerical capabilities. Based on a few test cases, we estimate $\mrt$ to vary by up to 20 \%, with some $Ra$ and $Pr$ combinations yielding a larger variance than other combinations. This could explain the exceptional case of $Ra=10^7$ and $Pr=100$, but the relatively small value of $\max(\mrt)=2.2$ in 2D could also be related to the shape and birth-frequency of plumes in this range of flow parameters (see Section IVc and Video S5 in our previous study).

Note also that unlike in Table II in our previous study, here we only provide one value for each $Ra$ and $Pr$, because there is no $\mrt$ splitting for the light vs.~heavy particle types, because $\St$ values are much smaller in this study (cf.~Fig.~\ref{fig:StL}).

\begin{table}[width=.9\linewidth,cols=4,pos=h]
\caption{Peak value of the normalized mean residence time, 2D}\label{tab2d}
\begin{tabular*}{\tblwidth}{@{} LLLL@{} }
\midrule
 Ra  & $Pr=1$ & $Pr=10$ & $Pr=100$ \\
\hline
  $10^5$ & >108 (65\%) &  >16 (65\%) & >25 (66\%) \\
  $10^7$ & >22 (73\%) & 3.4 & 2.2 \\
  $10^9$ & 2.1 & 6.3 & 2.5 \\
\hline
\bottomrule
\end{tabular*}
\end{table}

\begin{figure}[t]
    \centering
    \includegraphics[width=0.5\textwidth]{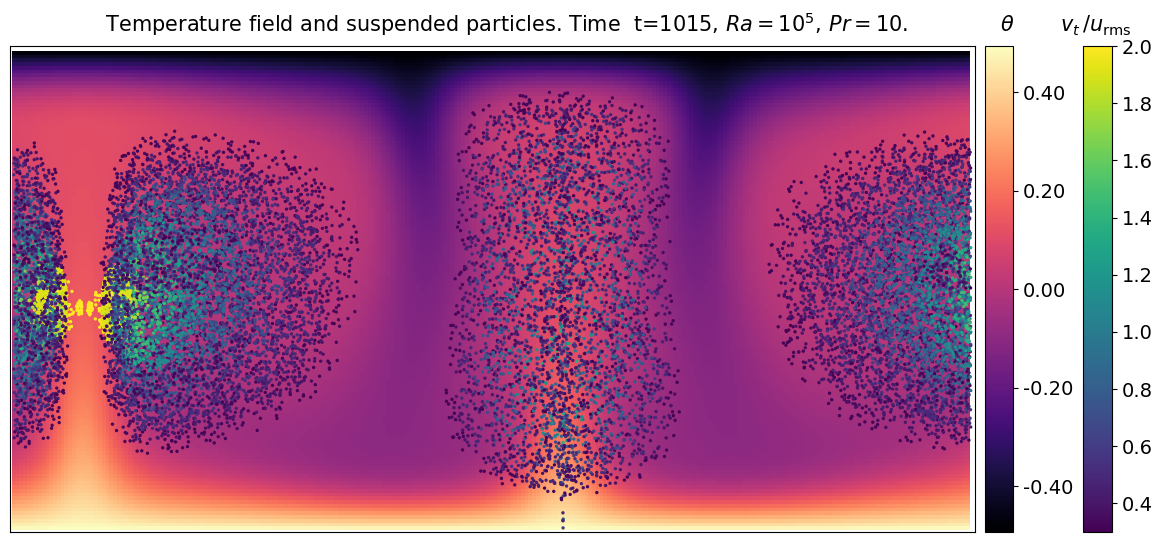}
    \caption{Particles that are still suspended at the time $t \approx 1000$ in 2D basally heated convection with $Ra=10^5, Pr=10$. Only the particle types satisfying $0.3 < \vt/\vm < 3.0$ are plotted.}
    \label{fig:c2d}
\end{figure}

In Fig.~\ref{fig:c2d}, we show a snapshot from the 2D simulation with $Ra=10^5$ and $Pr=10$. The convective cells force particles into regular trajectories. A similar effect is nicely illustrated for the case of cellular flow by \citet{Weinstein1988}, who show that the streamlines of $\bm{u}+\bm{v}_\mathrm{t}$ contain retention zones, in which the initial concentration of particles is maintained, while elsewhere in the flow particle concentration goes to zero \citep[see also the classical works of][]{Stommel1949,Maxey1987}. While the stronger upwelling in Fig.~\ref{fig:c2d} cuts the overlying particle cloud into two subdomains from which there is no escape in the plotted $\vt\,/\,\vm$ range, the less strong upwelling maintains a particle cloud with a time-decreasing concentration, because particles eventually fall down its conduit (see the dots near the root of the upwelling).

\begin{figure}
    \centering
    \includegraphics[width=0.48\textwidth]{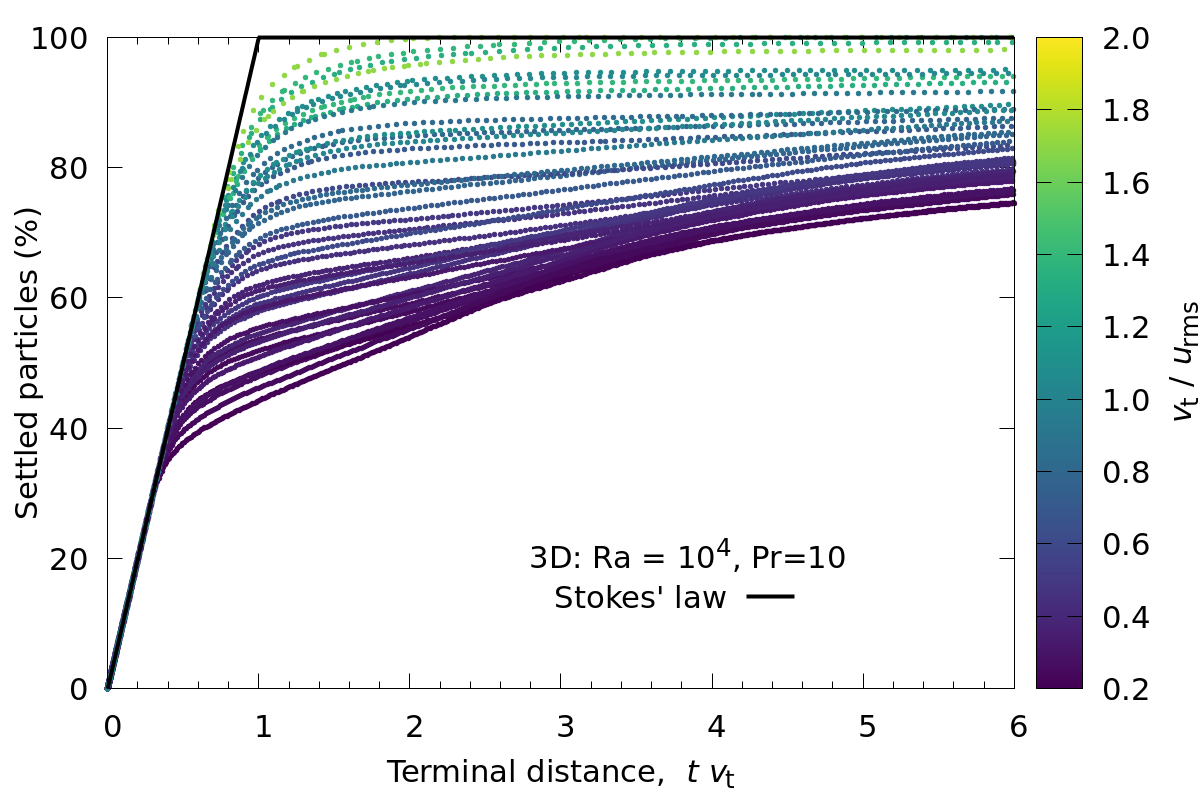}
    \includegraphics[width=0.48\textwidth]{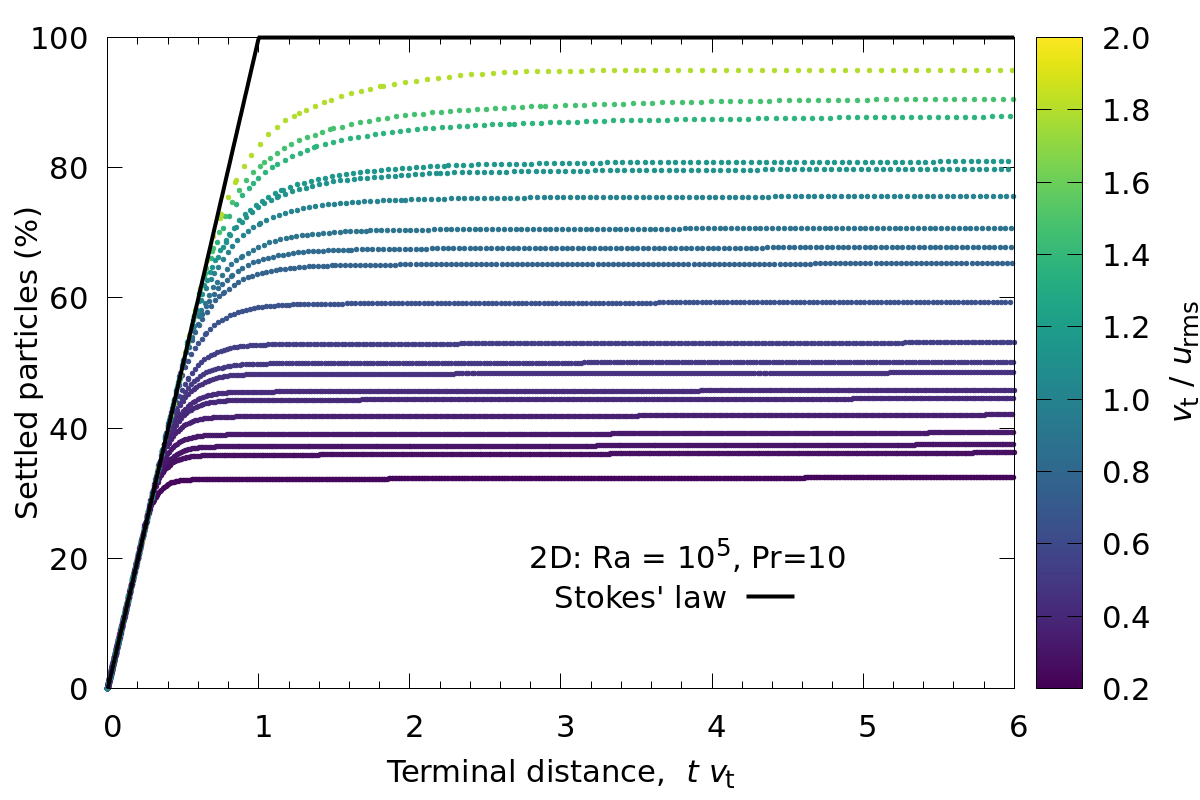}
    \caption{The percentage of settled particles as a function of time. Each line represents one of the 201 particle types, only those with $0.2<\avt/\vm<2.0$ are shown. The $x$-axis represents the terminal distance, $t\,\vt$, solid black line is the Stokes' law, i.e.~the solution for sinking in a quiescent fluid.}
    \label{fig:scurves}
\end{figure}

In Fig.~\ref{fig:scurves}, we compare the normalized settling curves for the 3D simulation with $Ra=10^4$ and $Pr=10$, studied in the main text, with those from the 2D convection depicted in Fig.~\ref{fig:c2d}. Each particle type is represented by one line, and only the types satisfying $0.02 \lesssim \avt/\vm \lesssim 2.0$ are plotted. The settling curves can typically be separated into two stages: in the first stage the fluid sweeps particles from regions that lie outside the retention zone, and the second stage describes the slow (or no) settling from the retention zone. The percentage of particles that settle in the first-stage is positively correlated with $\avt/\vm$, and the normalized settling rates become increasingly smaller in the second stage as $\avt/\vm$ is increased (cf.~also Fig.~\ref{fig:c3d}b from the main text).

Although the mean properties of the 3D flow analyzed in Fig.~\ref{fig:scurves} are also stationary ($Ra = 10^4$), the obtained settling curves are steeper in the second stage when compared to the completely flat settling curves obtained for the 2D convection, although its Rayleigh number is higher ($10^5$). It illustrates the effective imprisoning of particles in 2D stationary flows.

The mean residence times are larger in 2D also for non-stationary convection. Even when large-scale structures have time evolving shape and move horizontally within the model domain, 2D convective rolls effectively drag particle clouds along when out-of plane motion is prohibited. This is illustrated by the slow belt amplitude being still significant for $Ra=10^9$ and higher (see our previous study), while for the same flow parameters we do not observe settling retardation in 3D geometry.


\section{Internally heated convection}\label{sec:IH}

\begin{figure*}[t]
    \centering
    \includegraphics[width=1.0\textwidth]{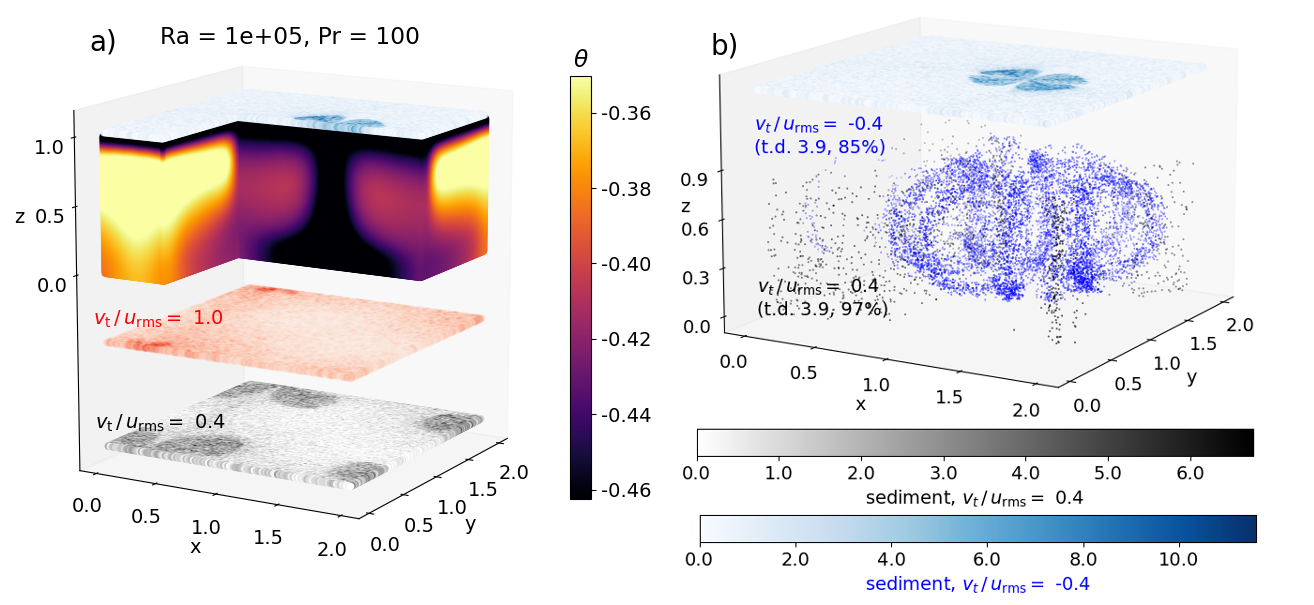}
    \caption{Same as panels a) and b) in Fig.~\ref{fig:c3d}, only here the convection is internally driven, with the equivalent Rayleigh number being $10^5$, and for a higher Prandtl number $Pr=100$. This time we show also the distribution of light particles with $\vt/\vm=-0.4$ (blue dots). These particles are focused in a two spherical clouds centred at conduits of the two dominant downwellings, and are deposited preferentially at their roots.}
    \label{fig:cIH}
\end{figure*}

In Fig.~\ref{fig:cIH}, we repeat the simulation with $Ra=10^5$ and $Pr=10$, only this time the heat is provided entirely via uniform internal heating instead of from below. That is, the bottom temperature boundary condition is no-flux instead of fixed value, and a spatially uniform heat production per unit mass $S$ is set such as to make the internal heating Rayleigh number, $Ra_\mathrm{IH} := \alpha g S H^5 /(\nu \kappa^2 c_p)$, equal to $10^5$. 

Due to active cooling from above, the downwellings are strong in the statistically steady state and dominate the flow, while upwellings serve merely as a return flow that is being pushed by the descent of cold material. As a result, the rising of light particles is strongly affected by the large-scale flow, because the light particles are prevented from reaching the top boundary by the two dominant downward currents. The settling of heavy particles, on the other hand, is much less inhibited by the passive return flow (Fig.~\ref{fig:cIH}b).

\begin{figure}
    \centering
    \includegraphics[width=0.48\textwidth]{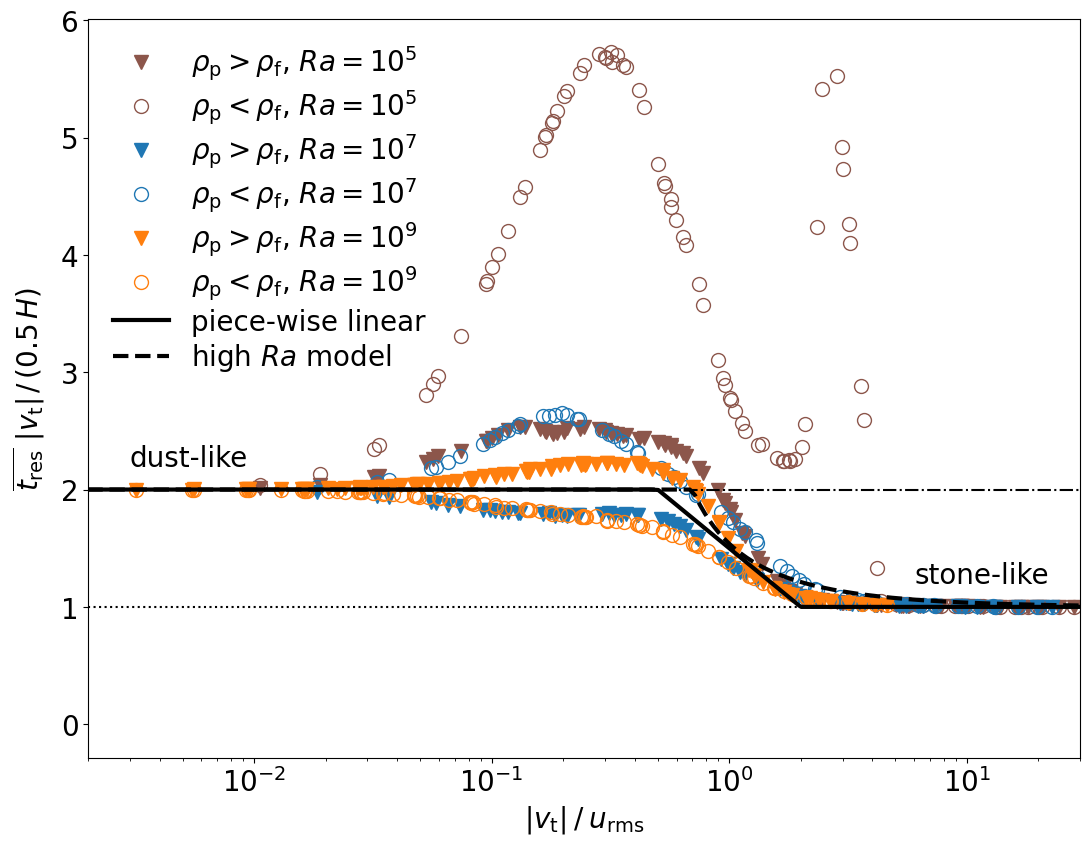}
    \caption{Same as Fig.~\ref{fig:ResT}, only here for internally instead of basally heated convection. Because of flow asymmetry, the residence times are different for the light and for the heavy particle types (empty circles vs.~filled triangles). As $Ra$ increases, $\mrt$ approaches the high-$Ra$ model from Eq.~\eqref{eqfinal}.}
    \label{fig:ResTIH}
\end{figure}

The asymmetry of the flow causes the mean residence time to differ for the light with respect to the heavy particle types (Fig.~\ref{fig:ResTIH}), but this $\mrt$ splitting is unrelated to preferential sampling, i.e.~to the centrifugal effect described in Section IV(B) of our previous study. Note also the local $\mrt$ maximum observed for $Ra=10^5$ and $Pr=10$ near $\avt \sim 3\vm$. We sometimes obtained such local peaks also for basally heated, $Ra=10^5$ convection, when the simulation was not in statistically steady state yet. Such local peaks can be caused by a thick current with a nearly vertical velocity $\bm{v}_\mathrm{curr}$ that is larger than the mean velocity of the flow, ${v}_\mathrm{curr} > \vm$, in which particles with $\bm{v}_\mathrm{t} \sim -\bm{v}_\mathrm{curr}$ nearly stay in place inside the current, locally increasing $\mrt$ near $\vt/v_\mathrm{curr}$ in effect (see e.g.~the yellow particles in Fig.~\ref{fig:c2d}).

As $Ra$ is increased, the mean residence time converges to Eq.~\eqref{eqfinal}, confirming the applicability of our high-$Ra$ model also to different convective modes. Note, however, that for the highest $Ra$ that we simulate ($10^9$) the flow asymmetry still generates a non-negligible light-heavy splitting of $\mrt$, and also a non-zero slow belt amplitude can still noticed. Confirming the robustness of our high-$Ra$ model is thus left to future numerical or analogue experiments with a yet higher $Ra$.

\bibliographystyle{cas-model2-names}

\bibliography{refs}


\end{document}